\documentclass[lettersize,journal]{IEEEtran}
 \pdfoutput=1

\usepackage{enumitem}
\usepackage[numbers,sort&compress]{natbib}
\usepackage{url}
\usepackage{amsmath,amssymb,amsfonts}
\usepackage{mathtools}
\usepackage{bm}
\usepackage{soul}
\usepackage{stmaryrd}
\usepackage{algorithmic}
\usepackage{graphicx}
\usepackage{multirow}
\usepackage{adjustbox}
\usepackage{subcaption}
\usepackage{textcomp}
\usepackage{xcolor}
\def\BibTeX{{\rm B\kern-.05em{\sc i\kern-.025em b}\kern-.08em
    T\kern-.1667em\lower.7ex\hbox{E}\kern-.125emX}}
\usepackage{amsthm}
\theoremstyle{definition}
\usepackage{xr-hyper}
\usepackage{hyperref}    
\usepackage{lipsum}
\usepackage{scalerel}
\usepackage{stackengine,wasysym}
\usepackage{cleveref}
\usepackage{sistyle}
\usepackage{arydshln}
\SIthousandsep{,}
\usepackage{setspace}

\input{mysymbol.sty}

\begin{document}

\title{Graph-based Algorithm Unfolding for Energy-aware Power Allocation in Wireless Networks

\author{
Boning Li$^\star$, 
Gunjan Verma$^\dag$, 
and Santiago Segarra$^\star$ \\
\textit{$^\star$Rice University, USA \hspace{1cm}  $^\dag$US DEVCOM Army Research Lab., USA}
}
\thanks{
Research was sponsored by the Army Research Office and was accomplished under Cooperative Agreement Number W911NF-19-2-0269. 
B. Li was partially supported by the Ken Kennedy Institute 2020/21 Ken Kennedy-Cray Graduate Fellowship.
The views and conclusions contained in this document are those of the authors and should not be interpreted as representing the official policies, either expressed or implied, of the Army Research Office or the U.S. Government. 
The U.S. Government is authorized to reproduce and distribute reprints for Government purposes notwithstanding any copyright notation herein.%
\newline
E-mails: \{boning.li, segarra\}@rice.edu, gunjan.verma.civ@army.mil.}}

\maketitle
\begin{abstract}
We develop a novel graph-based trainable framework to maximize the weighted sum energy efficiency (WSEE) for power allocation in wireless communication networks.
To address the non-convex nature of the problem, the proposed method consists of modular structures inspired by a classical iterative suboptimal approach and enhanced with learnable components. 
More precisely, we propose a deep unfolding of the successive concave approximation (SCA) method. 
In our unfolded SCA (USCA) framework, the originally preset parameters are now learnable via graph convolutional neural networks (GCNs) that directly exploit multi-user channel state information as the underlying graph adjacency matrix.
We show the permutation equivariance of the proposed architecture, {which is a desirable property for models applied to wireless network data.}
The USCA framework is trained through a stochastic gradient descent approach using a progressive training strategy. 
The unsupervised loss is carefully devised to feature the monotonic property of the objective under maximum power constraints. 
Comprehensive numerical results demonstrate its generalizability across different network topologies of varying size, density, and channel distribution. 
{Thorough comparisons illustrate the improved} performance and robustness of USCA over state-of-the-art benchmarks.
\end{abstract}

\begin{IEEEkeywords}
Wireless power allocation, multi-user multi-cell interference, weighted sum energy efficiency maximization, deep algorithm unfolding, graph convolutional neural networks. 
\end{IEEEkeywords}

\section{Introduction}\label{sec:intro}
\vspace{-1mm}
Current and future wireless communication networks involve {transmitting a massive amount of data} via their infrastructures while operating with limited resources~\cite{stanczak2009fundamentals,xu2021survey, chandrasekhar2009spectrum}.
Power is known to be a fundamental resource that requires efficient allocation to ensure the quality of wireless services~\cite{shannon1948mathematical}.
In the meantime, the increasing size of modern wireless networks requires sustainable energy management to fulfill user demands with minimal power consumption~\cite{chu2017path}.
\colored{Securing energy efficiency (EE) when allocating power for industrial wireless sensor networks can help reduce greenhouse gas emissions and electromagnetic pollution~\cite{calvillo2016energy,raj2022energy}.
For mobile devices, energy-efficient power allocation is vital to reduce electricity budgets and prolong their battery life}~\cite{yu2016fair}.
{Motivated by such urgent demands in real-world applications,} our work focuses on the energy-efficient allocation of transmit power in the uplink of a wireless interference network.\looseness=-1

Mathematically, optimal power allocation takes the form of maximizing a system-level EE performance function subject to constraints on resource budgets.
\colored{In interference-limited networks, typical EE metrics are non-convex in nature, including the system global EE (GEE), the minimum of node-wise EE (MEE), and the weighted sum of EE (WSEE).
Among them, WSEE is a particularly challenging objective -- taking the form of a sum of fractions with non-concave numerators, thereby rendering the WSEE maximization problem NP-hard~\cite{luo2008dynamic,li2013energy,salaun2018optimal,freund2001solving}.
}
Although monotonic programming has been employed to reduce the computational time of branch-and-bound procedures, the performance of such globally optimal methods scales poorly \colored{with the feasible space} and with the size of the network~\cite{matthiesen2018optimization}.
\colored{Hence, it is impractical for wireless deployment requiring real-time power allocation decisions, thus motivating several suboptimal techniques} to obtain approximate power policies while promoting scalability. 
\colored{Based on optimization models involved,} these classical methods can be broadly categorized into those based on interference cancellation~\cite{jin2020multiple,mai2022energy}, matching theory~\cite{fang2016energy,zhou2016energy}, Lagrangian dual decomposition~\cite{dong2014optimal,fang2019optimal}, and \colored{successive concave approximation} (SCA)~\cite{al2019energy,long2020non,zappone2017globally,yang2017unified,matthiesen2020globally,su2020energy,efrem2019dynamic,li2014max}.
Our special interest lies in the last of these classical approaches due to its popular adoption and {guaranteed first-order convergence~\cite{yang2017unified}. 
SCA has numerically proved near-optimal} performance in the maximization of GEE~\cite{al2019energy,long2020non,zappone2017globally} and MEE~\cite{li2014max,vilni2018energy}. 
However, for WSEE, even though SCA performs better than the other classical methods~\cite{nguyen2018energy}, there is still a notable gap in performance between SCA solutions and global optima even with only seven users~\cite{matthiesen2018optimization,matthiesen2020globally}.
In addition, the time-varying channel characteristics pose a harsher requirement on computational efficiency as the allocation needs to be faster than the channel coherence time. 
In this context, the nested iterative nature of SCA may preclude its practical implementation.
We seek to overcome these identified challenges by enhancing SCA with data-driven machine learning elements.

\IEEEpubidadjcol
Deep neural networks (DNNs) have shown promising success in wireless power allocation and other communication problems~\cite{rajapaksha2021deep,lee2019improving,sun2018learning,huang2022unsupervised,xu2019energy,lee2018deep,zhao2020power,eisen2019learning,eisen2020optimal,hou2021user,guo2021learning,sun2017learning,chowdhury2020unfolding,lin2020unsupervised,pellaco2020deep,kumar2021adaptive,zhao2021distributed,vanchien2020power}.
Most of these approaches are model-free, meaning that they parameterize some function of interest with established multi-purpose architectures such as multi-layer perceptrons (MLPs)~\cite{rajapaksha2021deep,lee2019improving,sun2018learning,huang2022unsupervised}, convolutional neural networks (CNNs)~\cite{xu2019energy,lee2018deep}, recurrent neural networks~\cite{zhao2020power}, and graph neural networks (GNNs)~\cite{eisen2019learning,eisen2020optimal,hou2021user}.
Hence, these approaches rely on the approximation capability of DNNs to mimic the optimal solutions of classical optimization techniques based on solved examples~\cite{csaji2001approximation}.
A significant limitation in supervised DNNs, as are adopted in many works above, is the requirement of a large set of labeled data, or optimally solved problem instances in our case, which may be infeasible for computationally intractable problems.
Another downside is that model-free DNNs do not tend to generalize well to unseen scenarios such as shifts in channel distributions, varying network sizes, or user mobility.
To that end, GNNs incorporating domain knowledge outperform their universal counterparts (as demonstrated in the setting of energy-agnostic sum-rate maximization~\cite{guo2021learning}), thus calling for specialized parameterization of power policy for energy-aware allocation as well.

Our proposed solution combines model-based and learning-based methodologies.
The proposed approach is learning-based in the sense that it leverages stochastic gradient descent (SGD) to learn GNN parameters from simulated channel data.
GNNs are able to exploit topological patterns of interference between users to locally process instantaneous channel state information (CSI).
At the same time, our method is model-based since it takes advantage of the classical SCA algorithm by informing a stacked neural network architecture that imitates the iterative structure of SCA.
Our goal is to achieve higher performance and faster computation than classical methods by combining the strengths of expert knowledge and statistical learning.
Motivated by this, we adopt the paradigm of algorithm unfolding~\cite{monga2019algorithm} to accomplish this synergistic combination.\looseness=-1

\vspace{1mm}
\noindent {\bf Related work.} 
\colored{While the application of deep learning to wireless power allocation is an active area of research, only a limited subset} of these works utilize graph-based learning methods~\cite{eisen2019learning,eisen2020optimal,chowdhury2020unfolding} or algorithm unfolding~\cite{sun2017learning,chowdhury2020unfolding,lin2020unsupervised, pellaco2020deep}.
The only prior work within power allocation that lives at the intersection (like our proposed approach)~\cite{chowdhury2020unfolding, chowdhury2021efficient} unfolds the iterative weighted minimum mean squared error (WMMSE) method for sum-rate maximization.
Nonetheless, sum-rate is a fundamentally different (not considering the efficiency of energy consumption) and theoretically simpler (being merely a special case of WSEE) objective than the WSEE here considered.\looseness=-1

On a broader note, algorithm unfolding was first developed as a fast approximation of an iterative sparse coding algorithm~\cite{gregor2010learning}.
Later on, many other iterative algorithms were unfolded, extending this scheme to a wide range of applications including image super-resolution~\cite{zhang2020deep}, multi-channel source separation~\cite{wisdom2016deep}, compressive sensing~\cite{zhang2020amp}, graph signal denoising~\cite{chen2020graph}, and precoding design~\cite{hu2020iterative}.
In this context, we propose the unfolded successive concave approximation (USCA) method. 
To the best of our knowledge, this is the first graph-based deep unfolded architecture for the iterative SCA algorithm. 
By achieving performance and computational gains out of graph-based learning and algorithm unfolding, {USCA yields state-of-the-art performance both in terms of the computational efficiency required to generate power allocations for wireless interference networks and the resulting energy efficiency of those power allocations.}

\vspace{1mm}
\noindent {\bf Contribution.} 
The main contributions of this work are:
\begin{itemize}[topsep=0pt, wide=0pt]
    \item[i)] {We propose a novel learning architecture, an unfolded version of SCA (USCA),} for energy-efficient power allocation in wireless interference networks.
    {Iterative solvers and updates in SCA are now} parameterized by learnable graph convolutional neural networks (GCNs).
    \item[ii)] {We provide a thorough analysis on the computational complexity of the proposed USCA along with guidelines for distributed implementation.} 
    Proposition~\ref{prop:1} shows that the proposed method is permutation equivariant as long as the augmented learnable components satisfy this property.
    \item[iii)] {
    The unique structure of USCA is designed to have an initialization with good performance by exploiting the topological information of the channel data.
    A novel regularization term enforcing monotonic penalty is introduced to incorporate domain knowledge, improving performance and robustness.}
    \item[iv)] Through numerical experimentation with wide-ranging settings and configurations, we exemplify the advantageous performance of the proposed method compared with state-of-the-art alternatives, its generalizability to variations in network size and channel distribution, {and the potential of transfer learning between different path-loss (PL) models.}
\end{itemize}

\vspace{1mm}
\noindent {\bf Paper outline.} 
In Section~\ref{sec:system}, we present the system model and formulate the power allocation problem as a constrained non-convex optimization problem. 
In Section~\ref{sec:method}, we briefly introduce the classical SCA algorithm and provide a detailed description of our proposed unfolded architecture. 
Theoretical analyses of permutation equivariance and computational complexity are presented in Sections~\ref{sec:method:perm} and~\ref{sec:method:comp}, respectively. 
In Section~\ref{sec:exp}, we illustrate results of thorough numerical experimentation, 
demonstrating the superior performance of USCA in comparison with the original SCA and other benchmark methods, as well as its generalizability to changes in network settings and channel distributions. 
Finally, Section~\ref{sec:conclusion} closes this paper with conclusions and future directions.

\vspace{1mm}
\noindent {\bf Notation.} 
Sets and maps are typeset in calligraphic letters such as $\ccalA$. 
Alternative notations for sets $\{a_i\}^n_{i=1}{\,=\,}\{a_1, ..., a_n\}$ and vectors $[a_i]^n_{i=1}{\,=\,}[a_1, ..., a_n]$ are occasionally used for the purpose of clarification; when clear from context, we omit the size.
Diagonal-related notations $\diag({\bbA})$ and $\diag({\bba})$ denote a diagonal matrix storing diagonal elements of matrix ${\bbA}$ or elements of vector ${\bba}$, respectively. 
${\bb0}$ and ${\bf 1}$ denote all-zeros and all-ones vectors of appropriate size.
Operations $|\cdot|$, $\|\cdot\|$, $(\cdot)^\top$, $(\cdot)^H$, $\mbE(\cdot)$, and $[\cdot]_+$ denote absolute value, $L^2$-norm, transpose, conjugate transpose, expected value, and positive part, respectively. 
For conditional statements, $\llbracket q \rrbracket{\,=\,}1$ if $q$ is true, 0 otherwise.
Elementwise multiplication between two vectors (or two matrices) is denoted as $\bba\odot\bbb$.
The clipping operation $\bbx{\coloneqq}\bbx|_{[a,b]}$ denotes $\bbx$ restricted to the set $[a,b]$.
When applied to vectors, scalar functions or operators are applied in an elementwise manner.
By default, $\log$ is in base 2.\looseness=-1


\vspace{-1em}
\section{System Model and Problem Formulation}\label{sec:system}
\vspace{-1mm}
We consider the uplink of a multi-cell interference network with $L$ single-antenna users and $M$ base stations (BSs). 
Each BS has $n_R$ antennas. 
Let $a(i)$ represent the BS serving user $i$ for $i{\,=\,}1,...,L$. 
Denoting the signal transmitted by $i$ as $x_i \in \mbC$, the received signal ${\bby}_{a(i)} \in \mbC^{n_R}$ at BS $a(i)$ is
\begin{equation}\label{eq:y}
    {\bby}_{a(i)} = \sum^L_{j=1}{\bbh}_{a(i),j}x_j + {\bbz}_{a(i)},
\end{equation} 
where ${\bbh}_{a(i),j}{\in}\mbC^{n_R}$ is the channel from $j$ to $a(i)$ and circularly-symmetric complex Gaussian noise ${\bbz}_{a(i)}{\sim}\ccalN_\ccalC(0, \sigma_i^2)$ is added at receiver $a(i)$ with power $\sigma_i^2$. 
Assuming matched filtering at the receiver {and perfect channel estimation}, the achievable data rate of the link from $i$ to $a(i)$ with bandwidth $B$ can be derived as
\begin{equation}\label{eq:r}
    R_i = B\log\left(1+\frac{\alpha_i p_i}{1+\sum_{j\neq i}\beta_{i,j} p_j}\right),
\end{equation} 
where the CSI terms
\begin{equation}
\alpha_i=\frac{\|{\bbh}_{a(i),i}\|^2}{\sigma_i^2} \quad \text{ and} \quad \beta_{i,j}=\frac{|{\bbh}^H_{a(i),i} {\bbh}_{a(i),j}|^2}{\sigma_i^2\|{\bbh}_{a(i),i}\|^2} \nonumber
\end{equation} 
are the channel gain over that link (which may be denoted by the user index for simplicity, i.e., link $i$) and the multi-user interference from $j$ to $i$, respectively. 
We define the multi-user CSI matrix  ${\bbH}\in \mbR^{L\times L}$ with channel gains as its diagonal entries and interference coefficients as off-diagonal entries, i.e. $H_{i,j}{\,=\,}\llbracket j{\,=\,}i\rrbracket\alpha_i + \llbracket j{\,\neq\,}i\rrbracket\beta_{i,j}$. 
Moreover, the average transmit power vector ${\bbp}{\,=\,}[p_i]^{L}_{i=1}$ is subject to maximum power constraints $\bm{P}_{m}{\,=\,}[P_{m,i}]^{L}_{i=1}$. 
Notice that by introducing the CSI matrix $\bbH$, the system can now be interpreted as a directed graph with users as its nodes and $\bbH$ as its adjacency matrix. 

Furthermore, the link's achievable rate over its power consumption defines its energy efficiency (EE) in bits per joule, i.e.,\looseness=-1
\begin{equation}\label{eq:ee}
    \EE_i = \frac{B\log\left(1+\frac{H_{i,i} p_i}{1+\sum_{j\neq i}H_{i,j} p_j}\right)}{\mu_i p_i+P_{c,i}},
\end{equation}
where $\mu_i$ is the amplifier inefficiency in the transmitter $i$, and $P_{c, i}$ is a constant portion of power consumed in $i$ unrelated to the communication.
With this notation in place, we formally state the maximum WSEE power allocation problem as follows\looseness=-1
\begin{equation}\label{eq:p1}\tag{P1}
\begin{aligned}
& \underset{{\bbp}}{\max}
& &\sum^L_{i=1}w_i\frac{\log\left(1+\frac{H_{i,i} p_i}{1+\sum_{j\neq i}H_{i,j} p_j}\right)}{\mu_i p_i+P_{c,i}} \\
& \text{s.t.}
& & 0 \leq p_i \leq P_{m,i}, \;\,\, i = 1, \ldots, L.
\end{aligned}
\end{equation}
As mentioned in Section~\ref{sec:intro}, among various EE metrics, we consider the specific case of WSEE as the objective in~\eqref{eq:p1}.
{In that objective, the constant bandwidth $B$ in the definition of EE [cf.~\eqref{eq:ee}] can be dropped as it does not affect the nature of this maximization problem.}
Since WSEE depends on {weighted efficiencies of individual users}, certain users can be easily prioritized over others, thus fostering flexibility in heterogeneous networks.
Nonetheless, it should be noted that the method proposed in Section~\ref{sec:method} is general enough to apply to other EE metrics (such as GEE and MEE) and additional optimization constraints.
For simplicity moving forward, in~\eqref{eq:p1} we consider uniform system constants $\mu_i{\,=\,}\mu$ and $P_{c,i}{\,=\,}P_{c}$ as well as uniform maximum power constraint $P_{m,i}{\,=\,}P_{m}$ for all users $i$. 
However, our method can be easily extended to the non-uniform case.\looseness=-1 

The objective in~\eqref{eq:p1} is a sum of fractions whose numerators are non-concave in general, making it an NP-hard problem~\cite{freund2001solving}. 
Several approaches have been proposed to approximately solve~\eqref{eq:p1}, but either {optimality}, tractability, or both have to be sacrificed~\cite{zappone2019wireless}. 
In practice, a few approximate methods aim to solve for a local maximum of~\eqref{eq:p1}. 
The performance of such traditional body of work depends on good initialization points, requiring the costly process of repeatedly solving a series of problems, each with a larger power constraint than its predecessor, until the actual target power constraint is reached. 
For this reason, we are motivated to develop an efficient and scalable method that can generate near-optimal power allocations in {near} real-time. 


\vspace{-1em}
\section{Unfolded Successive Pseudo-Concave Approximation (USCA) Framework}\label{sec:method}
\vspace{-1mm}
Deep algorithm unfolding~\cite{monga2019algorithm, balatsoukas2019deep, hershey2014deep} is {an algorithmic design paradigm} where iterative optimization algorithms are fused with neural networks for greater computation efficiency and/or prediction accuracy.
Zooming into the layers of unfolded neural networks, their architectures are usually inspired by existing algorithms specific to the task at hand. 
Since our system model, namely the multi-user wireless interference network, can be naturally modeled as a graph, we present our unfolded algorithm based on GNNs~\cite{kipf2017semi, gama2018convolutional, roddenberry2019hodgenet}. 
Indeed, we develop and evaluate a GCN-based framework~\cite{kipf2017semi} to unfold SCA, a first-order optimization algorithm for non-convex objectives.
To the best of our knowledge, this is the first graph-based unfolded algorithm for \emph{energy-efficient} power allocation.\looseness=-1 

In preparation for USCA, we first introduce the basics of SCA~\cite{yang2017unified,matthiesen2020globally}. 
Essentially, this classical algorithm handles the maximization of a non-concave function $f$ by maximizing a sequence of (pseudo-)concave surrogate functions $\{\Tilde{f}_t(\bbx;\bbx^{(t)})\}_{t{\,=\,}0}^T$.
These surrogate functions must satisfy a series of assumptions stated in~\cite[Section III]{yang2017unified}.
In the specific context of WSEE maximization, the objective in~\eqref{eq:p1} can be approximated through a series of concave functions (see \cite[Section IV]{matthiesen2020globally} for details)
\begin{align}\label{eq:weei}
    \begin{split}
        \sum\limits_{i=1}^L \widetilde{\EE}_i({\bbp};\bbp^{(t)}) 
        = \sum\limits_{i=1}^L c_1 R_i (p_i,\bbp_{-i}^{(t)} ) + c_2 p_i + c_3.
    \end{split}
\end{align}
In~\eqref{eq:weei}, $c_1$, $c_2$ and $c_3$ (whose explicit expression can be found in~\cite{matthiesen2020globally}) are constant coefficients given $\bbp^{(t)}$. 
{User $i$'s data rate $R_i(p_i,\bbp_{-i}^{(t)})$ [cf.~\eqref{eq:r}] is a function of its own power $p_i$ alone, while other users transmit at fixed powers $\bbp^{(t)}_{-i} = [p_j^{(t)}]_{j\neq i}^L$.}
We observe that $R_i(p_i,\bbp_{-i}^{(t)})$ is concave in $p_i$, from where the concavity of~\eqref{eq:weei} with respect to $p_i$ immediately follows.
Approximating the objective of~\eqref{eq:p1} by~\eqref{eq:weei}, we obtain the following concave maximization problem
\begin{equation}\label{eq:p2}\tag{P2}
\begin{aligned}
& \underset{{\bbp}}{\max}
& &\sum^L_{i=1} \widetilde{\EE}_i (\bbp; \bbp^{(t)}) \\
& \text{s.t.}
& & 0 \leq p_i \leq P_{m}, \;\,\, i = 1, \ldots, L.
\end{aligned}
\end{equation}
We sequentially solve~\eqref{eq:p2} following the update policy on $\bbp^{(t)}$ given by the line search
\begin{equation}\label{eq:update}
    \bbp^{(t)} = \bbp^{(t-1)} + \gamma^{(t)}  (\mbB\bbp^{(t)} - \bbp^{(t-1)}),
\end{equation}
where $\mbB\bbp^{(t)}$ is a global maximizer of~\eqref{eq:p2} at the $t$-th iteration.
Using the Armijo rule~\cite{armijo1966minimization}, a scalar step size $\gamma^{(t)}$ can be picked for a substantial update in the ascent direction, specific to the current iteration. 
Leveraging the convergence of the subproblems~\eqref{eq:p2}, it has been shown in~\cite{yang2017unified} that any limit point of $\{\bbp^{(t)}\}$ is guaranteed to be a stationary point of~\eqref{eq:p1}.
Moreover, each of the subproblems~\eqref{eq:p2} can be solved in polynomial time using standard convex optimization methods such as interior-point methods or projected gradient descent~\cite{ben1998robust,bubeck2014convex}. 
However, SCA requires nested iterations to solve~\eqref{eq:p2} and update~\eqref{eq:update} for convergence. 
Additionally,~\cite{matthiesen2020globally} has demonstrated that SCA depends on sophisticated initializations\footnote{When solving for $\bbp_K$ with constraint $P_{m_K}$, a sequence of $\{\bbp_0, ..., \bbp_K\}$ are sequentially solved for with ascending power constraints from $P_{m_0}$ through $P_{m_K}$. 
In each pass, the initial point for $\bbp_k$ is set to {the optimal solution obtained for } $\bbp_{k-1}$ (i.e., the optimal solution obtained for the previous smaller constraint).\label{ft:sca-init-p}}
to land in good local optima, which makes its computation even more inefficient. 

Intuitively, if we can find steeper {ascent} directions (i.e., a $\mbB\bbp^{(t)}$ in~\eqref{eq:update} that need not be the {exact global} maximizer of the concave approximation but can be closer to the true non-approximated {optimizer of the objective}) and better step sizes (i.e., self-adaptive step sizes that can be different in every dimension) at each iteration, we can achieve higher performance with fewer iterations of these enhanced SCA blocks. 
With this motivation, we elucidate in the following section our proposed unfolded architecture USCA, where key functions of SCA are augmented by learnable neural networks.

\vspace{-1em}
\subsection{Proposed unfolding framework}\label{sec:method:usca}
\begin{figure*}[t]
	\centering
    \includegraphics[width=.9\linewidth, trim=0 5cm 2cm 0, clip]{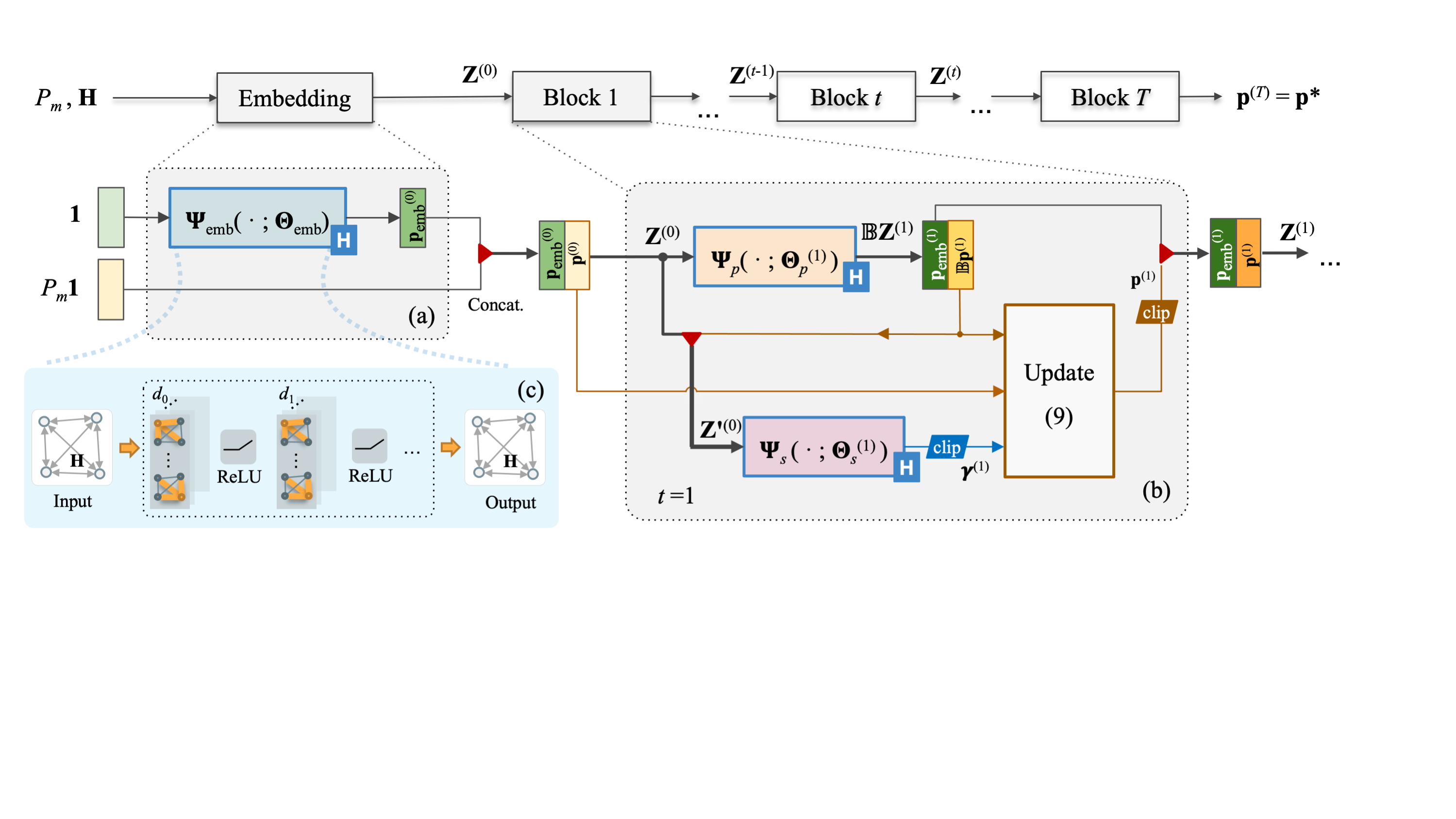}
    \vspace{-1mm}
    \caption{Overall schematic view of our proposed USCA architecture for energy-efficient power allocation. 
    (a)~The embedding block: 
    This block encodes channel information through a GCN $\Psi_{\emb}$, whose output is concatenated with $P_{m} \mathbf{1}$ to form the input to the first USCA block. 
    (b)~The USCA block: 
    This block contains two GCNs ($\Psi_p$ and $\Psi_s$) that respectively mimic the classical operations in SCA of solving problem~\eqref{eq:p2} and finding the step size to update the power allocation. 
    The outputs of these GCNs are combined following the update in~\eqref{eq:usca4}, whose functional form is inspired by the classical counterpart.
    (c)~The GCN: 
    A schematic view of a generic GCN with feature dimensions being $d_0$, $d_1, ...$ and ReLU non-linearities. The three GCNs in our model ($\Psi_{\emb}$, $\Psi_p$, and $\Psi_s$) employ this basic architecture.
    } \label{fig:usca}
\vspace{-2mm}
\end{figure*}

\vspace{-1mm}
\noindent{\bf Architecture.} 
Our unfolded architecture (depicted in Fig.~\ref{fig:usca}) is composed of multiple stacked layers where some key parameters, which used to be optimization-based, are now refined with learnable GCN-based subnetworks. 
More precisely, we propose to formulate the power allocation policy $\bbp{\,=\,}\Phi(P_m,\bbH; \boldsymbol{\Theta})$ as a function of the CSI matrix $\bbH$ and the power constraint $P_m$ through a layered architecture $\Phi$ with trainable weights $\boldsymbol{\Theta}{\,=\,}[{\bm \Theta_{\emb}}, {\bm \Theta_{p}^{(1)}}, {\bm \Theta_{s}^{(1)}}, ..., {\bm \Theta_{p}^{(T)}}, {\bm \Theta_{s}^{(T)}}]$. 
{To mimic the repetitive nature of SCA iterations, we enable parameter sharing across all layers, i.e., ${\bm \Theta_{p}^{(t)}}{\,=\,}{\bm \Theta_{p}}$ and ${\bm \Theta_{s}^{(t)}}{\,=\,}{\bm \Theta_{s}}, \forall t{\,=\,}1,...,T.$ 
With parameter sharing, one can extend the number of iterations without retraining, which is otherwise not possible because extra layers would introduce untrained parameters.
Moreover, unfolding models with shared parameters can be trained easier and faster, a point to be expatiated later in this section.
Terminologically, we may occasionally refer to a model with shared parameters as a {\it recurrent} model and the non-sharing one a {\it sequential} model. 
By default, our USCA takes the recurrent option; the sequential alternative is code-named with a suffix as \mbox{USCA-NS}.
Although parameter sharing is preferred and recurrent USCA is our primary choice, we still keep explicit superscripts on ${\bm \Theta_{p}^{(t)}}$ and ${\bm \Theta_{s}^{(t)}}$ in expressions to follow, to maintain the notation consistent with the most general case. 
}

To understand the USCA architecture in Fig.~\ref{fig:usca}, first notice that it consists of $T$ blocks, preceded by a single embedding block. The input to a generic block $t{+}1$ is $\bbZ^{(t)}{\,=\,}[\bbp_{\emb}^{(t)},\bbp^{(t)}]{\,\in\,}\reals^{L \times 2}$, which consists of the horizontal concatenation of an embedding vector $\bbp_{\emb}^{(t)}{\,\in\,}\reals^L$ and the current power allocation $\bbp^{(t)}{\,\in\,}\reals^L$ determined by the first $t$ blocks.
The power allocation output by USCA is $\bbp^{(T)}$, which can be directly read from the second column of $\bbZ^{(T)}$, the output of the $T$-th block.
The input $\bbZ^{(0)}$ to the first unfolded block is given by the embedding block in Fig.~\ref{fig:usca}. More precisely,
\begin{equation}\label{eq:z0}
    \bbZ^{(0)} = [\bbp_{\emb}^{(0)},\bbp^{(0)}] =  \left[ \Psi_{\emb}(\mathbf{1},\bbH; {\bm \Theta}_{\emb}), P_m \mathbf{1} \right],
\end{equation}
where $\bbp^{(0)}$ is set as a constant vector of value $P_m$ and $\bbp_{\emb}^{(0)}$ is obtained as the output of the GCN $\Psi_{\emb}$.\looseness=-1 

For completeness, we provide here the structure of a generic $Q$-layer GCN $\Psi$ with input $\bbZ$ and trainable parameters $\bbTheta$ acting on the CSI matrix $\bbH$.
Denoting the input as $\bbX^0{\,=\,}\bbZ$, then $\bbZ'{\,=\,}\Psi(\bbZ,\bbH; \bm{\Theta}){\,=\,}\bbX^Q$, where an intermediate $q$-th layer of the GCN is given by
\begin{equation}\label{E:gcn}
	\mathbf{X}^{q} = \sigma_q \left(
	\bbD^{-\frac{1}{2}}\bbH\bbD^{-\frac{1}{2}}\mathbf{X}^{q-1}\bm{\Theta}^{q}\right).
\end{equation}
In~\eqref{E:gcn}, $\bbD{\,=\,}\diag(\bbH \mathbf{1})$, ${\bbTheta}^{q}{\,\in\,}\mathbb{R}^{d_{q-1}{\times}d_{q}}$ are the trainable parameters, $d_{q-1}$ and $d_{q}$ are the dimensions of the output features of layers $q{-}1$ and $q$, respectively, and $\sigma_q(\cdot)$ is the activation function. 
The specifics of the number of layers $Q$, the hidden dimensions $d_q$, and the activations functions $\sigma_q$ used in our experiments are further detailed in Section~\ref{sec:exp}.

We are left to explain the inner workings of a generic unfolded block $t$, so that we can obtain $\bbZ^{(t)}$ from $\bbZ^{(t-1)}$ (detailed in Fig.~\ref{fig:usca} for $t{\,=\,}1$). 
The first operation is given by the GCN $\Psi_p$, from which we obtain
\begin{equation}\label{eq:usca2}
	\mbB\bbZ^{(t)} = [\bbp_{\emb}^{(t)},\mbB\bbp^{(t)}] = \Psi_p(\bbZ^{(t-1)},\bbH; {\bm \Theta}_p^{(t)}).
\end{equation}
The first column $\bbp_{\emb}^{(t)}$ of the output in~\eqref{eq:usca2} updates the node embeddings and directly constitutes the first column of {$\bbZ^{(t)}$}.
On the other hand, the second column $\mbB\bbp^{(t)}$ of the output in~\eqref{eq:usca2} is used to update $\bbp^{(t)}$ following the expression [cf.~\eqref{eq:update}]
\begin{equation}\label{eq:usca4}
    \bbp^{(t)} = \left( \bbp^{(t-1)} + {\bbgamma}^{(t)}\odot (\mbB\bbp^{(t)} - \bbp^{(t-1)}) \right) \bigg\rvert_{[0,P_m]}.
\end{equation}
There are several differences between~\eqref{eq:usca4} and its classical counterpart~\eqref{eq:update}.
First, in~\eqref{eq:usca4} we obtain $\mbB\bbp^{(t)}$ as the output of a GCN $\Psi_p$ compared to the more computationally complex procedure of solving~\eqref{eq:p2} in classical SCA.
Second, the output of the update in~\eqref{eq:usca4} is clipped to the interval $[0,P_m]$ to guarantee that $\bbp^{(t)}$ is a feasible power allocation for any intermediate block $t$.
Lastly, the step size ${\bbgamma}^{(t)}$ in~\eqref{eq:usca4} is a vector (as opposed to a scalar in~\eqref{eq:update}) and is set via a GCN $\Psi_s$ instead of the Armijo rule. More precisely, we have that
\begin{equation}\label{eq:usca3}
    \boldsymbol{\gamma}^{(t)} = \left( \Psi_s([\bbZ^{(t-1)},\mbB\bbp^{(t)}],\bbH; {\bm \Theta}_s^{(t)}) \right)\bigg\rvert_{[0,1]}.
\end{equation}
The step size expression in~\eqref{eq:usca3} is motivated by the classical line search in that it depends on both the previous $\bbp^{(t-1)}$ (contained in $\bbZ^{(t-1)}$) and the solution $\mbB\bbp^{(t)}$ to the current approximation. 
Besides,~\eqref{eq:usca3} extends the classical Armijo rule to accommodate different step sizes for different nodes, instead of a common one for every node. 
In addition, the channel information is also taken into consideration in the step size computation. 
Intuitively, this additional flexibility can help in achieving faster convergence.

The entire framework is trained by minimizing a tailored loss function and following a {progressive strategy}, which we elaborate in the sequel. 

\vspace{1mm}
\noindent{\bf Loss function.} 
Given parameters $\boldsymbol{\Theta}$, the allocated power from USCA for a channel-constraint instance, {i.e., a channel instance $\bbH$ associated with the constraint $P_m$,} is $\bbp{\,=\,}\Phi( P_m, \bbH; \bm{\Theta})$. 
By denoting the objective of~\eqref{eq:p1} as $\WSEE(\bbp, \bbH)$, where we have made explicit its dependence on the power allocation and the channel state, we define the unsupervised loss function
\begin{equation}\label{eq:lossu}
        \ccalL_u(\bm{\Theta})=-\mbE_{\bf{H}\sim\ccalH} \left[ \WSEE(\Phi( P_m, \bbH; \bm{\Theta}), \bbH) \right],
\end{equation} 
where $\ccalH$ is a known channel distribution from which we draw samples of $\bbH$. 

We augment this loss with two regularizers that allow us to incorporate additional domain knowledge to our learning procedure.
The first regularizer leverages the fact that the WSEE objective in~\eqref{eq:p1} must be monotonically non-decreasing with respect to the maximum allowable power $P_m$ in the constraints.
The intuition is straightforward that larger values of $P_m$ correspond to strictly larger feasible sets in~\eqref{eq:p1}, which can never result in smaller objective values.
Accordingly, we devise a regularizer for our USCA loss that penalizes deviation from this desired monotonic relation.
More precisely, we set a small $\Delta P{\,>\,}0$ and for every $P_m$ of interest we consider the output of our USCA model when perturbing the maximum allowable power as ${\bbp}_{-}{\,=\,}\Phi\left( P_m{\,-\,}\Delta P, \bbH; \bm{\Theta}\right)$.
The monotonic regularization is then defined as 
\begin{equation}\label{eq:mono}
    \ccalR_m(\bm{\Theta}) =\ccalR_1(\bm{\Theta}) + \lambda_s \ccalR_2(\bm{\Theta}),
\end{equation}
where $\ccalR_1{\,=\,}[\text{WSEE}({\bbp}_{-},\bbH){\,-\,}\text{WSEE}({\bbp},\bbH)]_+$ is a hinge loss that penalizes violations from monotonicity and 
$\ccalR_2$ is a self-supervisory Huber loss {between the predictions for the original and perturbed problems} for instances with positive $\ccalR_1$, i.e., $\ccalR_2(\bm{\Theta}){\,=\,}\llbracket \ccalR_1{\,>\,}0 \rrbracket\,\phi(\bar{\bbp}, \bar{\bbp}_-)$, where $\phi$ is the classical Huber loss for robust regression~\cite{huber1992robust}.
The relative importance between $\ccalR_1$ and $\ccalR_2$ is given by the coefficient $\lambda_s$ (default: \num{1e3}). 
Notice that $\ccalR_m(\bm{\Theta})$ is non-zero only when monotonicity is violated.

As a second regularizer we consider a supervised loss whenever solved instances of~\eqref{eq:p1} are available.
The supervision can then be expressed as 
\begin{equation}\label{eq:sup_loss}
    \ccalL_s(\bm{\Theta}) = \phi(\Phi( P_m, \bbH; \bm{\Theta}), \bbp_\opt),
\end{equation}
where $\phi$ is a Huber loss and $\bbp_\opt$ is a pre-specified optimal power allocation for channel $\bbH$ and power constraint $P_m$.
Nevertheless, it must be stressed that the training of USCA does not necessarily require supervision since we have the option of not incorporating the supervised loss in~\eqref{eq:sup_loss}.
In fact, global optimal allocations can be hard to acquire, as we emphasized in Section~\ref{sec:system}. 
That being the case, we identify USCA as an unsupervised method since the access to optimal power allocations $\bbp_\opt$ is optional rather than essential. 

To sum up, we propose the following overall loss function as the weighted sum of all the aforementioned terms [cf. \eqref{eq:lossu}, \eqref{eq:mono}, and~\eqref{eq:sup_loss}]
\begin{equation}\label{eq:loss}
    \ccalL(\bm{\Theta}) = \ccalL_u(\bm{\Theta}) + \eta_m\ccalR_m(\bm{\Theta}) + \eta_s\ccalL_s(\bm{\Theta}),
\end{equation}
where $\eta_m$ and $\eta_s$ (default: 0) are configured to balance the main loss with {monotonicity} and supervision, respectively.
Leveraging the fact that $\ccalL(\bm{\Theta})$ in~\eqref{eq:loss} is differentiable with respect to $\bm{\Theta}$, we seek to minimize the loss via SGD.
Additional details on the training strategy are provided next.\looseness=-1

\begin{figure}[t] 
\centering
    \begin{subfigure}{\columnwidth} 
    \centering
        \includegraphics[width=\columnwidth
        ]{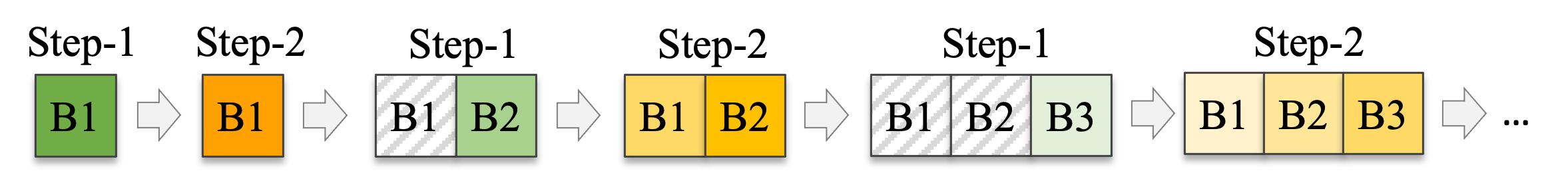}
        \vspace{-4mm}
        \caption{Training sequential models w/o parameter sharing.        
        }
        \label{fig:prog_train:nws}
    \end{subfigure}\hfil
    \begin{subfigure}{\columnwidth}
    \centering
        \includegraphics[width=\columnwidth,trim=0 2mm 0 0,clip,
        ]{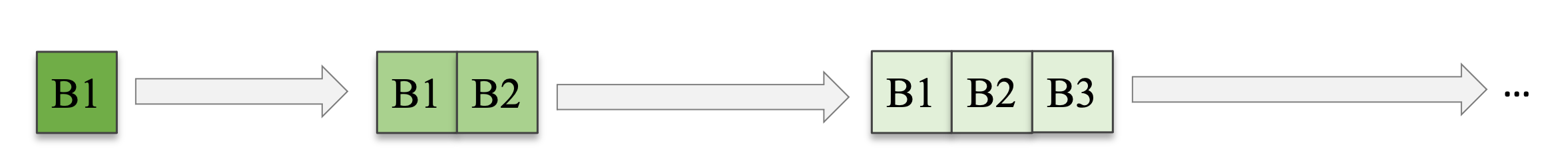}
        \vspace{-4mm}
        \caption{Training recurrent models with parameter sharing.
        }
        \label{fig:prog_train:ws}
    \end{subfigure}
\vspace{-1mm}
\caption{
{Conceptual illustrations of progressive training steps.}
}\label{fig:prog_train}
\vspace{-2mm}
\end{figure}

\vspace{1mm}
\noindent{\bf Training strategy.} In order to stabilize training, especially for deeper USCA architectures (larger number of blocks $T$), we use a progressive training strategy to enhance our end-to-end learning process. 
Similar strategies are widely used for {both sequential and recurrent models in the deep unfolding literature~\cite{wu2019sparse, chen2018theoretical,vu2020progressive}. }
In principle, the idea is to train the blocks one-by-one while decaying the learning rate by depth.
{General steps are illustrated in Fig.~\ref{fig:prog_train}.
Although we prefer to use recurrent USCA, we first describe the more sophisticated sequential training strategy for USCA-NS, which can easily be reduced to the recurrent strategy. }

{For sequential models (Fig.~\ref{fig:prog_train:nws}), we begin with training} the parameters of the $t{\,=\,}1$ block (in our case, the embedding block as well). 
We then freeze the trained parameters and train the next $t{\,=\,}2$ block. 
After this, we fine-tune all parameters within the trained blocks.
Subsequently, all previous blocks are fixed and we proceed to add the $t{\,=\,}3$ block into the training procedure, and so on. 
As training progresses to deeper blocks, learning rates of shallower blocks are exponentially decayed (the learning rate for the $t$-th block is $l_t{\,=\,}l_0d^{\,t-1}$) to prevent their parameters from varying too wildly.
In Fig.~\ref{fig:prog_train:nws}, different colors distinguish the two steps, namely 1)~freezing previous blocks while training the current block and 2)~fine-tuning all blocks, where lighter colors indicate smaller learning rates. 
For recurrent models, we simply combine the two steps because all blocks, though indexed differently, are essentially sharing the same set of parameters.
Thus, it does not make sense to fix certain blocks or to have learning rates decaying block-wise.
Rather, all blocks of USCA should coincide throughout the learning steps, as depicted in Fig.~\ref{fig:prog_train:ws}.
Upon $t$ blocks, all parameters are trained with a unified learning rate of $l_t{\,=\,}l_0d^{\,t-1}$.\looseness=-1

The proposed training method should be interpreted differently from classical layered neural networks where different layers tend to learn distinct functionalities~\cite{zeiler2014visualizing}. 
On the contrary, all USCA blocks are producing the same type of output in essence; all provide a feasible power allocation vector.
{In the sequential or non-parameter-sharing case}, if the previous $t{-}1$ blocks have been well trained and their output $\bbp^{(t{-}1)}$ (which is also the input to the $t$-th block) is close-to-optimal, training the \mbox{$t$-th} block would become much easier -- something not too far away from an identity map.
{Alternately, in the recurrent or parameter-sharing case, these blocks directly} share functional similarity and can indeed benefit from progressive training.
Unless otherwise stated, all unfolding models reported in Section~\ref{sec:exp} are trained using the above strategy.

\vspace{-1em}
\subsection{Permutation equivariance}\label{sec:method:perm}
\vspace{-1mm}
As one of the most popular implementations of graph-based neural networks, GCNs are simple and concise, making them a suitable choice for our $\Psi$ functions in USCA.
GCNs are composed of graph filters which are similar to convolutional filters in classical CNNs~\cite{segarra2017optimal}.
Those graph filters learn to perform appropriate local aggregations through trainable parameters.
Importantly, selecting GCNs as our functions $\Psi$ confers our USCA framework with {\it permutation equivariance}, as we explain next.
Let us begin by formally introducing this property.
Consider the set $\ccalF$ of functions $f{:\,}\reals^{N{\times}N}{\,\to\,}\reals^{N{\times}M}$, the set $\ccalG$ of functions $g{:\,}\reals^{N{\times}M'}{\,\times\,}\reals^{N{\times}N}{\,\to\,}\reals^{N{\times}M}$, and a generic permutation matrix $\bm{\Pi}{\,\in\,}\{0,1\}^{N{\times}N}$.

\definition{\it 
A function $f{\,\in\,}\ccalF$ is permutation equivariant if $f(\bm{\Pi} \bbH \bm{\Pi}^\top){\,=\,}\bm{\Pi} f(\bbH)$ for all matrices $\bbH$ and all permutations $\bm{\Pi}$. Similarly, a function $g{\,\in\,}\ccalG$ is permutation equivariant if $g(\bm{\Pi} \bbZ, \bm{\Pi} \bbH \bm{\Pi}^\top){\,=\,}\bm{\Pi} g( \bbZ, \bbH)$ for all matrices $\bbZ, \bbH$ and all permutations $\bm{\Pi}$.}\label{def:1}
\vspace{2mm}

With permutation equivariance, simply changing the ordering or labeling of nodes in a network at the input does not change the output values in each individual node. 
Rather, the outputs are reordered or relabeled in the same manner as the input.
We emphasize this property because it is particularly germane to our application where node (or user) indexing is arbitrary and, thus, should not cause any bias to power allocation results.
Notice that the entire USCA framework can be interpreted as belonging to $\ccalF$ for $N{\,=\,}L$ and $M{\,=\,}1$ since it maps $\bbH$ to $\bbp$. By contrast, individual functions $\Psi$ within USCA admit as inputs $\bbH$ and a matrix of node features, thus belonging to $\ccalG$.
Indeed, as building blocks of USCA, the functions $\Psi$ determine whether or not the whole framework is permutation equivariant, as we show next.

\proposition{\it If all basic subnetworks $\Psi_\emb$, $\Psi_p$, and $\Psi_s$ are permutation equivariant, then the USCA framework $\Phi$ is also permutation equivariant.} \label{prop:1} 

\begin{proof} 
We want to show that $\Phi(P_m,\bbPi\bbH\bbPi^\top; \boldsymbol{\Theta}){\,=\,}\bbPi\Phi(P_m,\bbH; \boldsymbol{\Theta})$ for every $\bbH$ and $\bbPi$, where the parameters $\boldsymbol{\Theta}$ and the maximum power $P_m$ are fixed.
This can be proved by showing that the embedding block and a generic USCA block $t$ in Fig.~\ref{fig:usca} are permutation equivariant, and by noting that composing a series of these operations preserves permutation equivariance.

That the embedding block is permutation equivariant directly follows from $\Psi_\emb$ satisfying this property combined with the fact that the second column of the block's output $\bbp^{(0)}$ is a constant vector $P_m \mathbf{1}$. Thus, $\bbp^{(0)}{\,=\,}\bbPi \bbp^{(0)}$ for any permutation matrix $\bbPi$.

For a generic USCA block $t$, we want to show that {when its input is} $\tilde{\bbZ}^{(t-1)}{\,:=\,}\bbPi \bbZ^{(t-1)}$ {with the channel matrix being} $\tilde{\bbH}{\,:=\,}\bbPi \bbH \bbPi^\top$, the output of the block is given by $\tilde{\bbZ}^{(t)}{\,=\,}\bbPi \bbZ^{(t)}$. 
Notice that we use $(\tilde{\cdot})$ to denote intermediate variables within the USCA block when the input and the channel matrix have been permuted.
From~\eqref{eq:usca2} and permutation equivariance of $\Psi_p$, it follows that $\tilde{\bbp}_{\emb}^{(t)}{\,=\,}\bbPi {\bbp}_{\emb}^{(t)}$ and $\mbB\tilde{\bbp}^{(t)}{\,=\,}\bbPi \mbB\bbp^{(t)}$. Moreover, combining this last equality with~\eqref{eq:usca3} and recalling that $\Psi_s$ is assumed to be permutation equivariant we have that\looseness=-1
\begin{align}
    \tilde{\boldsymbol{\gamma}}^{(t)} &= \left( \Psi_s([\tilde{\bbZ}^{(t-1)},\mbB\tilde{\bbp}^{(t)}],\tilde{\bbH}; {\bm \Theta}_s^{(t)}) \right)\bigg\rvert_{[0,1]} \nonumber \\
    & = \left( \Psi_s(\bbPi[{\bbZ}^{(t-1)},\mbB{\bbp}^{(t)}],\bbPi{\bbH}\bbPi^\top; {\bm \Theta}_s^{(t)}) \right)\bigg\rvert_{[0,1]}\!\! \!\! = \! \bbPi {\boldsymbol{\gamma}}^{(t)}. \nonumber
\end{align}
Finally, inputting these expressions into the update~\eqref{eq:usca4} we get that
\begin{align}
    & \tilde{\bbp}^{(t)} = \left( \tilde{\bbp}^{(t-1)} + {\tilde{\bbgamma}}^{(t)}\odot (\mbB\tilde{\bbp}^{(t)} - \tilde{\bbp}^{(t-1)}) \right) \! \bigg\rvert_{[0,P_m]} \nonumber \\ 
    & \!\!  = \!\left( \bbPi {\bbp}^{(t-1)} + { \bbPi {\bbgamma}}^{(t)} \!\odot \! ( \bbPi \mbB{\bbp}^{(t)} - \bbPi {\bbp}^{(t-1)}) \right) \! \bigg\rvert_{[0,P_m]} \!\!\!\! = \bbPi {\bbp}^{(t)}, \nonumber
\end{align}
where the last equation follows from the fact that all the operations in the update are elementwise.
Hence, we have established that the output of a generic USCA block is given by $\tilde{\bbZ}^{(t)}{\,=\,}[ \tilde{\bbp}_{\emb}^{(t)}, \tilde{\bbp}^{(t)}]{\,=\,}[\bbPi \bbp_{\emb}^{(t)}, \bbPi \bbp^{(t)}]{\,=\,}\bbPi \bbZ^{(t)}$, as wanted.
\end{proof}

Combining Proposition~\ref{prop:1} with the well-established fact that GCNs are permutation equivariant~\cite{gama2020stability}, we can conclude that \emph{the proposed USCA framework is permutation equivariant}. 

Conveniently, the modular structure of USCA depicted in Fig.~\ref{fig:usca} enables the practitioner to choose alternative architectures for the subnetworks $\Psi$.
The loss function can also be seamlessly reformulated if different objective functions or constraints are of more interest.
In Section~\ref{sec:exp}, the performance of other candidate $\Psi$ architectures is introduced and compared. 
In this setting, Proposition~\ref{prop:1} guarantees that, as long as the $\Psi$ functions are permutation equivariant, the USCA framework will also possess this favorable property.

\begin{remark}[Flexibility to varying number of users]\normalfont
Different from many neural network applications with fixed input and output dimensions, communication systems are a more challenging scenario as the number of users in the system can vary over time.
This requires USCA to be able to output a power allocation for a number of users $L'$ different from the $L$ users with which it was trained.
This distinguishing capability is achieved by relying on graph neural networks and is numerically showcased in Section~\ref{sec:exp:network}.
\end{remark}

\vspace{-1em}
\subsection{Computational complexity}\label{sec:method:comp}
\vspace{-1mm}
In the broad field of non-convex optimization, a fundamental benefit of learning-based methods over traditional model-based algorithms lies in the reduced computational complexity when presented with a new sample.
In contrast to model-based methods which have to solve the optimization problem~\eqref{eq:p2} from scratch 
whenever an unseen sample is given, learning-based methods present a fundamental trade-off where expensive computations for training (e.g., computing gradients and back propagation, and constructing the dataset if required) can be performed {\it offline}, leaving only  simple computations  (e.g., affine transformation, scalar multiplication, and pointwise nonlinearity) to be performed {\it online} regularly. 
Retraining or fine-tuning may be recommended when the real input distribution shifts far away from the trained one, but only sporadically and, typically, with fewer data.

In terms of the offline complexity, it should be recalled that USCA can be trained in the absence of solved instances of the power allocation problem.
{Hence, USCA effectively circumvents the need to solve several NP-hard problems for training.
Even though the offline training may still cost several minutes to hours (depending on the size of dataset),}
nevertheless, since we are interested in the feasibility of real-time implementation, we {concentrate on the {\it online} computational complexity in the following}.\looseness=-1

To better understand the online complexity of USCA, let us focus on a generic unfolding block in Fig.~\ref{fig:usca} since the embedding block has fewer computations and equivalent asymptotic complexity. 
From~\eqref{E:gcn} it follows that a generic USCA block involves $\sum_{q=1}^{Q}(L^2 d_{q-1} + d_{q-1} L d_{q})$ real multiplications (additions neglected) and $\sum_{q=1}^{Q+1} L d_{q}$ non-linear activations to compute $\mbB\bbZ^{(t)}$ as the output of $\Psi_p$. 
Similarly, computations in the other GCN subnetwork $\Psi_s$ can be broken down as above. 
Together with clipping operations and updating rule~\eqref{eq:usca4} -- each requiring $\ccalO(L)$ complexity --, the overall time complexity of USCA with respect to the number of users is $\ccalO(L^2)$.
On the other hand, classical SCA requires several solutions of the convex optimization problem~\eqref{eq:p2} each of them with polynomial complexity (whose specific form will depend on the algorithm chosen) plus determining the step size $\gamma(t)$ with linear time complexity. 
As illustrated in Section~\ref{sec:exp:compare}, it is not only the case that one USCA block entails less computations than solving one instance of~\eqref{eq:p2} but also the number of USCA blocks $T$ needed for good performance is typically less than the iterations needed in classical SCA.
On top of that, the progressive training strategy can further minimize the number of blocks in the final USCA framework by simply terminating at the block where EE metrics flatten out.

Lastly, USCA is amenable to a distributed deployment, which can bring about extra time efficiency in practice.
This feature is inherited from the graph neural networks $\Psi$ underlying USCA, which can be naturally implemented in a distributed manner~\cite{segarra2017optimal}.
However, it should be noted that training should be performed in a centralized manner and, once training is complete, the whole set of USCA parameters ${\bm \Theta}$ must be broadcast to every user.
Although the offline phase cannot be distributed, a distributed online phase can facilitate scalability of the proposed solution to large communication systems.

\vspace{-1em}
\section{Numerical Evaluation}\label{sec:exp}
\vspace{-1mm}
{We demonstrate the robust and state-of-the-art performance of USCA} through extensive numerical experiments in various settings\footnote{The implementation of our method is available at \url{https://github.com/bl166/usca_power_control}. All experiments were conducted using an NVIDIA GeForce GTX 1080 Ti Graphics Card.}.
{Section~\ref{sec:exp:compare} validates the superior performance of USCA compared with other approaches}, where training and test data are drawn from the same channel distribution. 
We then {evince the generalizability} of the proposed method to changes in the network density and size (Section~\ref{sec:exp:network}) and to changes in the channel distribution (Section~\ref{sec:exp:dist}).
Finally, in Section~\ref{sec:exp:supervision} we analyze up to what extent fine-tuning on a limited number of samples drawn from a distribution $\ccalH_1$ different from the training distribution $\ccalH_0$ can aid in transfer learning on $\ccalH_1$.\looseness=-1

Our experimental setup follows the one considered in~\cite{matthiesen2020globally}.
More precisely, we consider $M$ $n_R$-antenna BSs located at the center of $M$ ($M{\,=\,}m^2$ for some $m{\,\in\,}\mbZ$) adjacent non-overlapping square cells of side $1 \km$ forming a square overall coverage area of side $m\km$.
In each BS, the received noise power $\sigma^2{\,=\,}FN_0B$ is defined with noise figure $F{\,=\,}3\dB$, noise density $N_0{\,=\,}{-}174\dBmHz$, and bandwidth $B{\,=\,}180\kHz$.
Within the overall coverage area, $L$ single-antenna users, all with $P_c{\,=\,}1\Watt$ and $\mu{\,=\,}4$, are placed {at random locations}.
We primarily consider $M{\,=\,}4$, $L{\,=\,}8$, $n_R{\,=\,}1$, as well as the wideband spatial {(WBS)} path loss (PL) propagation model~\cite{calcev2007wideband} with Rayleigh fast fading for the generation of training data. 
Test data may be configured differently and this will be explicitly discussed as we go into details about configurations of each experiment. 
Unless otherwise specified, the training dataset includes {$4000$} random channel realizations {split into training and validation subsets in a 4-fold} cross validation (CV) manner. 
{Each channel realization is then} coupled with \num{51} maximum power constraints which are integers in decibel units ranging from $-40$ to $10\dBW$,
{i.e., $P_m{\,\in\,}\ccalP_m{\,=\,} \{p{\,\in\,}\mbZ\,|\,{-}40{\,\leq\,}p{\,\leq\,}10\dBW\}$.
In total, this gives us \num{204000} channel-constraint samples. }
The test set has {\num{51000} channel-constraint samples, constituted by another $1000$} independent channel realizations and identical power constraints as the training set. 

\begin{figure}[t]
	\centering    \includegraphics[width=.98\linewidth
    ]{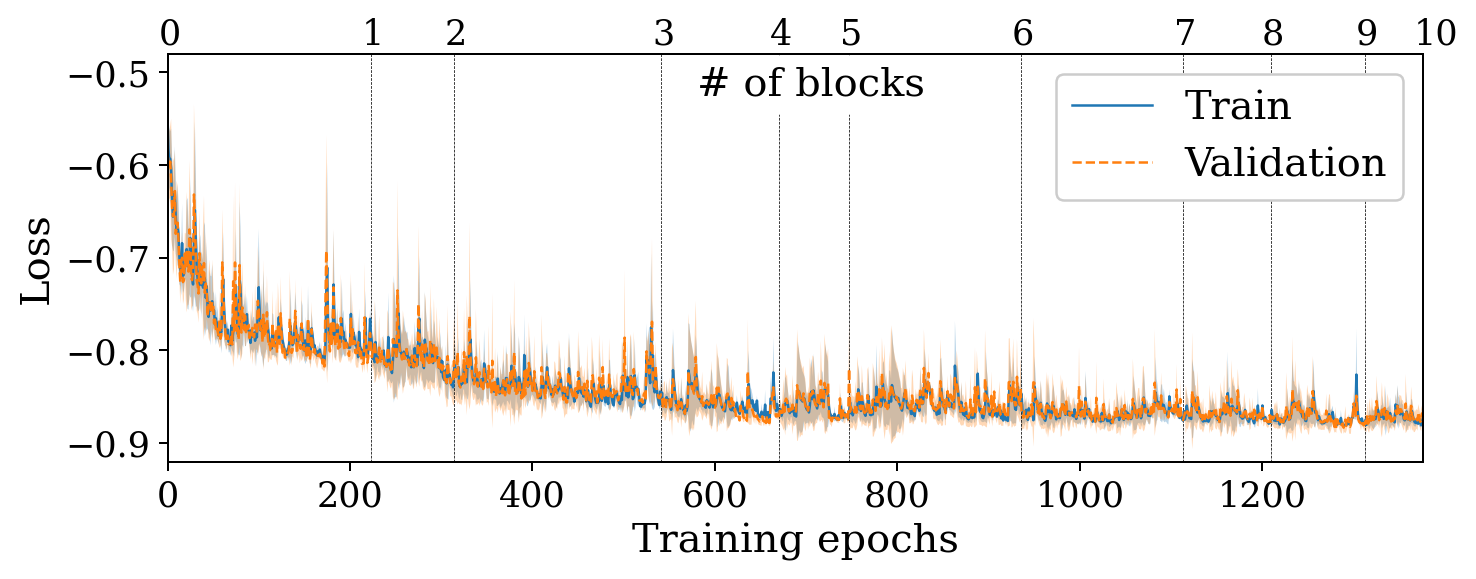}
    \vspace{-2mm}
    \caption{{
    Learning curves of \mbox{GCN-USCA} showing the training and validation losses against the number of completed epochs.
    The termination of each progressive training milestone is annotated on the top}.\looseness=-1
    } \label{fig:learn-curve}
    \vspace{-4mm}
\end{figure}

{Next, we elaborate on the hyperparameters which are grid searched and configured using the set that produces the best validation performance for each model to test.}
Our proposed USCA method solves~\eqref{eq:p1} through {11 stacked} blocks (1 embedding block {followed by} $T{\,=\,}10$ unfolding blocks).
{This number is picked by growing the number of blocks in the model until the performance stops increasing. 
Unfolding blocks $t{\,=\,}2$ through $T$ contain replicated parameters of the $t{\,=\,}1$ block, following the recurrent design.}
The subnetworks $\Psi$ are given by 5-layer GCNs with hidden dimensions equal to {$\{16, 64, 64, 64, 16\}$ with the ReLU nonlinearity.
Adam optimizer~\cite{kingma2014adam} with a base learning rate of $l_0{\,=\,}5{\times}10^{-4}$ (with a decay factor of $d{\,=\,}0.6$) and ${\ell}_2$ coefficient of $1{\times}10^{-6}$ is employed.
Fig.~\ref{fig:learn-curve} demonstrates the convergence of \mbox{GCN-USCA} throughout its progressive training process.
The shaded curves show the averages and standard deviations of all CV folds.
Each block is trained for a maximum of \num{1000} epochs with early stopping triggered by over 50 epochs of unchanged best WSEE performance on the validation set to avoid over-training. 
Another prevention of over-fitting is to adopt a $0.5$ dropout rate during training.
Due to early stopping, the number of training epochs for each block may be fold-dependent. 
For a visualization purpose, the plotted curves in Fig.~\ref{fig:learn-curve} continue until two out of the four folds are terminated.
At the beginning of each epoch, all training samples are randomly shuffled and split into non-overlapping mini-batches of size \num{2040}.
} 

\vspace{-1em}
\subsection{Performance comparison}\label{sec:exp:compare}
\vspace{-1mm}
To begin with, we present a thorough comparison between the performance of our proposed GCN-based USCA algorithm (\mbox{GCN-USCA}, or simply USCA) with that of the original SCA algorithm and several other established {benchmarks}. 
{A list of candidate methods is detailed} as follows. 
\begin{enumerate}[wide]
    \item {\bf SCA} is the original algorithm inspiring this work and is also a commonly used algorithm to address {power allocation problems}. 
    We set a sufficiently large upper limit of 500 inner {concave solver} iterations and 1000 outer iterations of successive approximation.
    Respectively, inner iterations solve a single instance of~\eqref{eq:p2} via projected gradient descent, and outer iterations fulfill updates as described in~\eqref{eq:update}.
    \item {\bf \mbox{Tr-SCA}}, or the truncated SCA, {institutes the achievable WSEE} of SCA with as many iterations as USCA. 
    {Inner and outer iteration numbers of \mbox{Tr-SCA} are sharply reduced to 5 and 10}, respectively, to match the  {configuration of} USCA (5 {GCN layers} in $\Psi$ and $T{\,=\,}10$ unfolding blocks). 
    {As with SCA, it also takes the intricate steps of power initialization} as mentioned in~\cref{ft:sca-init-p} (\Cpageref{ft:sca-init-p}).
    \item {\bf \mbox{MLP-USCA}} is implemented as an MLP-based USCA framework, simply by replacing GCNs $\Psi$ with MLPs, {in order to show the versatility of graph-based learning principles in wireless communication networks. }
    The input vector to each block is a concatenation of flattened logarithmic values of $\bbH$ and intermediate features $\bbZ^{(t)}$. 
    {Accordingly, ELU activation is used instead of ReLU to prevent dead nodes due to the possibly negative values in the inputs}.\looseness=-1
    \item {\bf GCN} without unfolding is included to highlight the effectiveness brought about by the SCA-inspired architecture. 
    This is a plain GCN with {$\bbH$ as the adjacency matrix and, at node~$i$, $P_m$ as the input signal and $p_i$ as the output signal.}
    To compensate for the capacity of learnable parameters, it is implemented as a 5-layer GCN with hidden dimensions {expanded to $\{32, 128, 128, 64, 32\}$} and trained in an end-to-end manner with $\eta_m{\,=\,}0.25$; 
    \item {\bf \mbox{Max-Pow}} designates to all users the maximal possible  transmit power, i.e., $p_i{\,=\,}P_m$ for all $i{\,=\,}1, ... , L$.
    This naive strategy works well for small $P_m$ values before interference starts to dominate EE metrics.
    \item {{\bf Opt} is the globally optimal solution given by the branch-and-bound procedure using monotonic optimization~\cite{matthiesen2020globally}.
    Due to its high complexity for large network sizes, similar to~\cite{matthiesen2020globally}, we provide Opt results only for the 6-user network.}
    {
    \item {\bf \mbox{GCN-USCA-NS}} is the sequential (or non-parameter-sharing) version of \mbox{GCN-USCA}, in which
    we do {\it not} enforce that ${\bm \Theta_{p}^{(t)}}{\,=\,}{\bm \Theta_{p}^{(t^\prime)}}$ or ${\bm \Theta_{s}^{(t)}}{\,=\,}{\bm \Theta_{s}^{(t^\prime)}}$ for any $t{\,\neq\,}t^{\prime}$.
    {Each of its blocks has identical numbers of hidden layers and nodes as the recurrent GCN-USCA. 
    In other words, it contains about $T$ times more trainable parameters than a GCN-USCA model of the same length}.\looseness=-1
    }
\end{enumerate}
\vspace{-1mm}

\begin{figure*}[t] 
\centering
    \begin{subfigure}{0.32\linewidth}
        \includegraphics[width=\linewidth,trim=0 6mm 0 0,
        ]{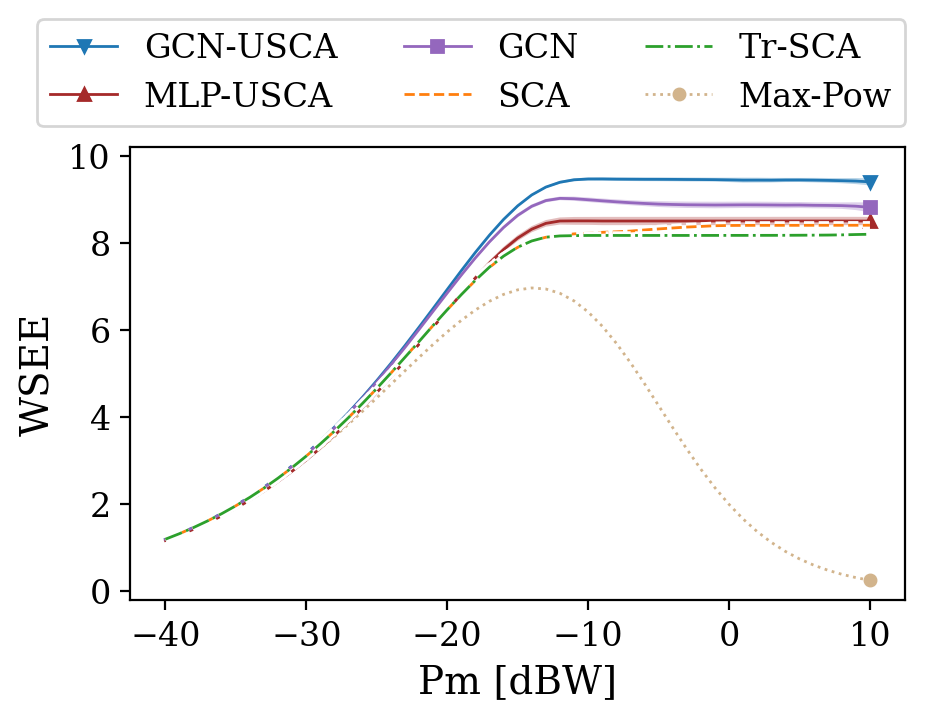}
        \vspace{-1em}
        \caption{
        }
        \label{fig:wsee:pmax}
    \end{subfigure}\hfil
    \begin{subfigure}{0.32\linewidth} 
        \includegraphics[width=\linewidth,trim=0 6mm 0 0,
        ]{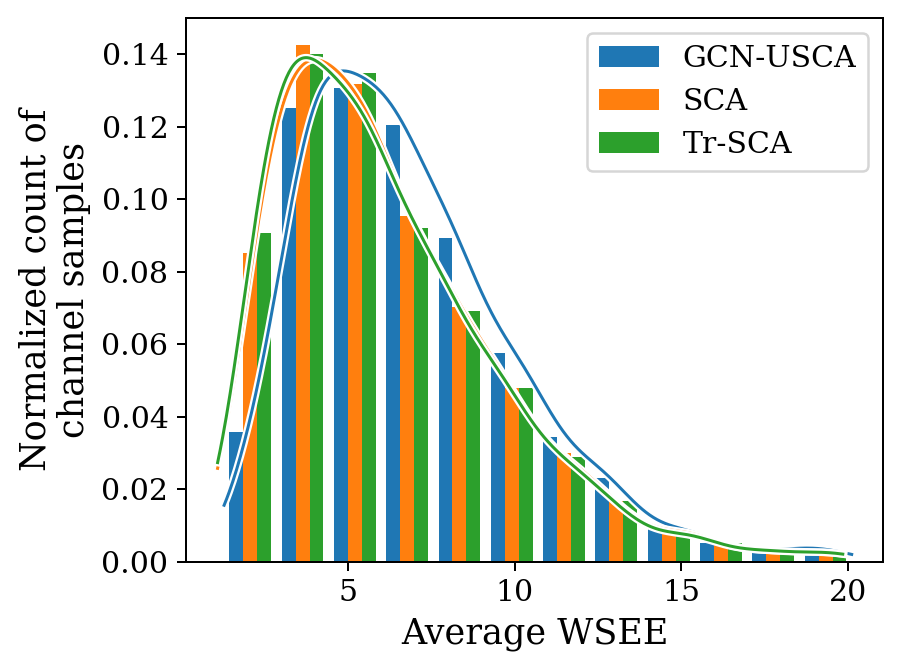}
        \vspace{-1em}
        \caption{
        }
        \label{fig:wsee:hist}
    \end{subfigure}\hfil
    \begin{subfigure}{0.32\linewidth}
        \includegraphics[width=\linewidth,trim=0 6mm 0 0,
        ]{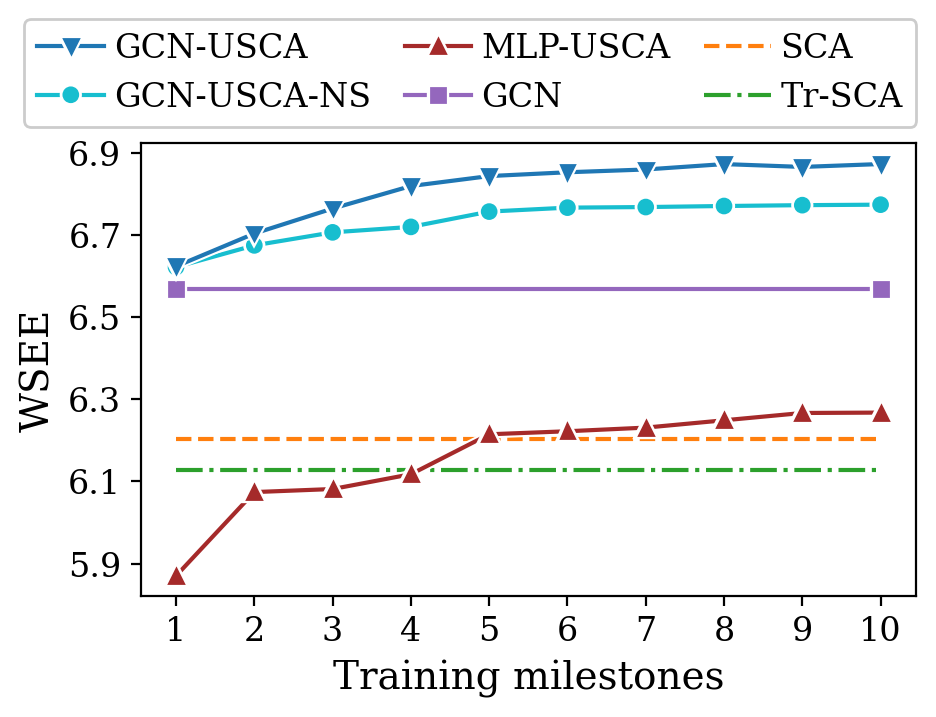}
        \vspace{-1em}
        \caption{
        }
        \label{fig:wsee:seq}
    \end{subfigure}
\vspace{-2mm}
\caption{
Performance comparisons of the proposed GCN-based USCA with SCA, a truncated version thereof, an MLP-based USCA, a vanilla GCN, and a naive policy that always allocates the maximum power $P_m$. 
(a)~Performance on the test set compared with all baselines.
Every point in these curves is given by the average WSEE over all test channel samples under the corresponding power constraint value.
(b)~Histogram of the achieved WSEE values for test channels compared with model-based competitors.  
Each data point represents the average WSEE value over all constraint values of a single channel sample.
(c)~Performance at progressive training milestones compared with learning-based competitors. 
The average WSEE values over all test channels and all constraint values are plotted against the number of unfolding blocks that have been trained and included in the framework. 
}\label{fig:wsee}
\vspace{-4mm}
\end{figure*}

We perform these comparisons on the basis of {the WBS Rayleigh fading model} and report the results in Fig.~\ref{fig:wsee}.
In a nutshell, the proposed \mbox{GCN-USCA} outperforms all other methods. 
{Fig.~\ref{fig:wsee:pmax} exhibits the average WSEE values achieved by candidate methods against \colored{different choices of $P_m$ values 
[cf.~\eqref{eq:p1}]} over 1000 test channel realizations.
The result of \mbox{GCN-USCA-NS} is elided here for a more visible view of the favorable \mbox{GCN-USCA} (the performance of \mbox{GCN-USCA-NS} is later presented in Tables~\ref{tab:wsee} and~\ref{tab:dist}).}
From the figure it follows that \mbox{GCN-USCA} yields the highest WSEE values for every choice of the power constraint bound $P_m$.
In particular, \mbox{GCN-USCA} largely {improves WSEE} with respect to the SCA baseline, underscoring the value of the proposed unfolding paradigm.
Moreover, when comparing \mbox{GCN-USCA} with other learning-based methods, namely \mbox{MLP-USCA} and GCN, it is clear that our proposed \mbox{GCN-USCA} performs better than both, especially in the high $P_m$ regime.
The WSEE values reached by \mbox{GCN-USCA}, \mbox{MLP-USCA} and GCN plateau at $P_m{\,\approx\,}{-}10\dBW$ with overall averages of 6.865, 6.269, and 6.581$\bpJ$, respectively.
The fact that \mbox{GCN-USCA} works better than both GCN and \mbox{MLP-USCA} indicates that its performance boost is attained by the combination of the unfolding architecture and the use of graph neural networks.

In Fig.~\ref{fig:wsee:hist}, we further compare \mbox{GCN-USCA} with the classical SCA and its truncated version {Tr-SCA}.
Notice that due to randomness in channel generation, the achievable {WSEE} of each channel realization can be notably different even with fixed power constraint.
{This is manifested by the histograms in which a point signifies the average WSEE} over all power constraints for a channel realization.
According to these empirical distributions, \mbox{Tr-SCA} matches the performance of SCA for most channel realizations (even though being surpassed in a few cases), but they both {are significantly beaten} by \mbox{GCN-USCA}. 
To be exact, \mbox{GCN-USCA}, SCA, and \mbox{Tr-SCA} respectively achieve
average WSEE values of {6.865, 6.203, and 6.127}$\bpJ$.
Compared with the full SCA, the performance degradation in \mbox{Tr-SCA} is not surprising because of the limited number of iterations allowed.
\mbox{Tr-SCA} pays a price in allocation suboptimality for lower computational cost.
{In terms of the wall-clock time}, allocating power under the 51 constraint values for one channel instance costs on average {$0.28$ seconds for \mbox{Tr-SCA} but $5.23$ seconds for SCA.
In contrast, \mbox{GCN-USCA} prevails by learning a better commencing power,} more {accurate} approximations, and {more directional updates} to converge faster to the {optimal solution} within as many iterations as \mbox{Tr-SCA}.
It takes only $0.09$ seconds to finish one pass of the aforementioned task.  
Additionally, \mbox{GCN-USCA} {works} in a one-shot manner, i.e., it treats individual channel-constraint samples as independent instances. 
This is fundamentally different from and favored over the classical benchmarks where allocation under higher constraint values needs to be {initiated through adequately} small constraint values (e.g., {$P_m{\,\ll\,}{-}20\dBW$, observable from} Fig.~\ref{fig:wsee:pmax}) to ensure good performance.\looseness=-1

{In Fig.~\ref{fig:wsee:seq}, the unfolding methods, namely \mbox{MLP-USCA} and \mbox{GCN-USCA} {(as well as its non-parameter-sharing implementation \mbox{GCN-USCA-NS}), are analyzed in depth.}
Their} allocation performance (average WSEE values over the entire test set) is compared progressively at $10$ training milestones.
The $t$-th milestone is defined as the {partially trained} framework at the time point exactly before proceeding to the \mbox{$(t{+}1)$-th} block.
Although the simple GCN does not have milestones defined, we draw a reference line at the level of its final performance.
{The obvious gains of \mbox{GCN-USCA} over \mbox{GCN} highlight the effectiveness of leveraging the unfolding paradigm.
Plus, the overwhelming advantage of \mbox{GCN-USCA} over \mbox{MLP-USCA} emphasizes the significance of exploiting the network topology.
Delving into more detailed comparisons, we} show the importance of utilizing the {topology embedding, the parameter sharing schematic, and the progressive training strategy from three aspects:}
i)~\mbox{GCN-USCA} reaches {higher performance than} the naive GCN at the {very first} milestone, which shows that our framework, {starting with a topology embedding block} and following a sophisticated unfolding architecture, can achieve higher performance with {significantly} fewer parameters.
ii)~\mbox{GCN-USCA} takes a lead over \mbox{GCN-USCA-NS} after their identical performance at the 1st milestone, which shows that parameter sharing is suitable for unfolding, for it preserves the repetitive processes and better mimics the iterative structures of SCA.
iii)~After the {8th} milestone, {it is visually bespoken that the average} performance of \mbox{GCN-USCA} cannot be further improved by adding more blocks, which indicates that we can safely cut it off at {fewer} blocks, further reducing the computational cost.  
Beware that the practical choice of $T$ should depend on the desired trade-off between optimality and computational budget.
For GCN-USCA, this is a very flexible choice owing to its recurrent nature. 
We confirm (up to 100 iterations) that redundant iterations will not negatively affect the final performance of GCN-USCA.

Finally, we discuss how prior knowledge or domain expertise can help design the learning-based framework.   
Recall that in proposing~\eqref{eq:mono} we reasoned that the achieved WSEE should be monotonically non-decreasing with the maximum admissible power $P_m$.
From Fig.~\ref{fig:wsee:pmax} it is evident that all methods (except for the naive \mbox{Max-Pow}), even those learning-based in which this knowledge is not hard coded, {generally} obey this monotonic pattern imposed by our carefully crafted regularizer.
Especially, in the GCN experiment, we observed significant improvement of tuning $\eta_m$ and eventually choose the empirical value of $0.25$. 
This concrete example validates the importance of utilizing domain knowledge to train data-driven models for desirable behaviors.

\vspace{-1em}
\subsection{Scalability against variations in network size}\label{sec:exp:network}
\vspace{-1mm}
Our previous results are focused on a test set with $M{\,=\,}4$ BSs and $L{\,=\,}8$ users. 
To demonstrate the scalability of the proposed method, we now showcase its performance with different network sizes.
In particular, the scenario of varying $L$ is of special relevance since, due to mobility, users might join and leave the system at any time.
We also consider the case of varying $M$ to assess whether a model trained on a smaller communication infrastructure can scale to a larger system.
More specifically, we generate new channels with $(L,\!M){\,\in\,}\{(6,\!4),$ $(12,\!4), (18,\!9), (48,\!16), (100,\!16)\}$.
With such settings, we can fix either the capacity $M$ or density $L/M$ (average number of users served by a BS) of the network while investigating how robust our proposed method is against changes in the network size $L$.
The models being tested are directly taken from Section~\ref{sec:exp:compare} (thus, trained on $M{\,=\,}4$ and $L{\,=\,}8$), and their performance is shown in Table~\ref{tab:wsee}.\looseness=-1

\renewcommand{\tabcolsep}{5pt}
\begin{table}[t]
    \caption{{
    Average WSEE values of power allocations on test sets with different number of users associated with different number of BSs. The best performance for every scenario is in bold font and the true optimum (when computationally feasible) is underlined.}}\label{tab:wsee}
    \vspace{-1mm}
    \begin{adjustbox}{max width=\columnwidth}
    \centering
    \begin{tabular}{r|c c c|c|c c}
        \hline
         \#BS ($M$)& \multicolumn{3}{c|}{4} & \multicolumn{1}{c|}{9} & \multicolumn{2}{c}{16}\\\hline
        \#Users ($L$)& 6 & 8     & 12    & 18     & 48     & 100\\\hline
        GCN-USCA	& {\bf 5.962} & {\bf 6.865} & {\bf 9.066} & {\bf 15.742} & {\bf 33.153} & {\bf 38.823} \\
        GCN-USCA-NS	& 5.959 & 6.830 & 8.968 & 15.664 & 32.267 & 34.633 \\
        MLP-USCA	& - & 6.269 & - & - & - & - \\
        GCN	& 5.789 & 6.581 & 8.583 & 15.105 & 31.130 & 36.358 \\
        SCA	& 5.731 & 6.203 & 7.429 & 13.435 & 25.780 & 30.000 \\
        Tr-SCA	& 5.655 & 6.127 & 7.354 & 13.404 & 25.755 & 29.978 \\   
        Opt & {\underline{6.087}} & - & - & - & - & - \\
        \hline
    \end{tabular}
    \end{adjustbox}
    \vspace{-4mm}
\end{table}

The first thing to notice is that, excluding the method computing the true optimum {for $M{\,=\,}4$ and $L{\,=\,}6$, our proposed architecture \mbox{GCN-USCA} always performs the best.
For that exception,} the system is small enough for us to compute the true optimal power allocation.
Thus, we can see that \mbox{GCN-USCA} reduces in $65\%$ [from $0.356\,({=\,}6.087{-}5.731)$ to $0.125\,({=\,}6.087{-}5.962)$] the suboptimality gap of classical SCA while being computationally more efficient as explained in Section~\ref{sec:exp:compare}.
Also, we can observe that the performance gain of \mbox{GCN-USCA} over \mbox{GCN-USCA-NS} grows with the network size (exceptionally apparent in 100-user networks), which implies better generalizability with parameter sharing.
Lastly, note that \mbox{MLP-USCA} can only be tested in the same setting on which it was trained since its input and output dimensions are fixed.
This further underscores the importance of considering GCNs for the subnetworks $\Psi$ in the unfolded architecture.
GCNs not only lead to better performance within the configuration seen during training but also enable the generalization to systems of varying size, a necessary requirement in real-world deployment.

\vspace{-1em}
\subsection{Robustness to path-loss model mismatches}\label{sec:exp:dist}
\begin{figure}[t]
	\centering
    \includegraphics[width=\linewidth]{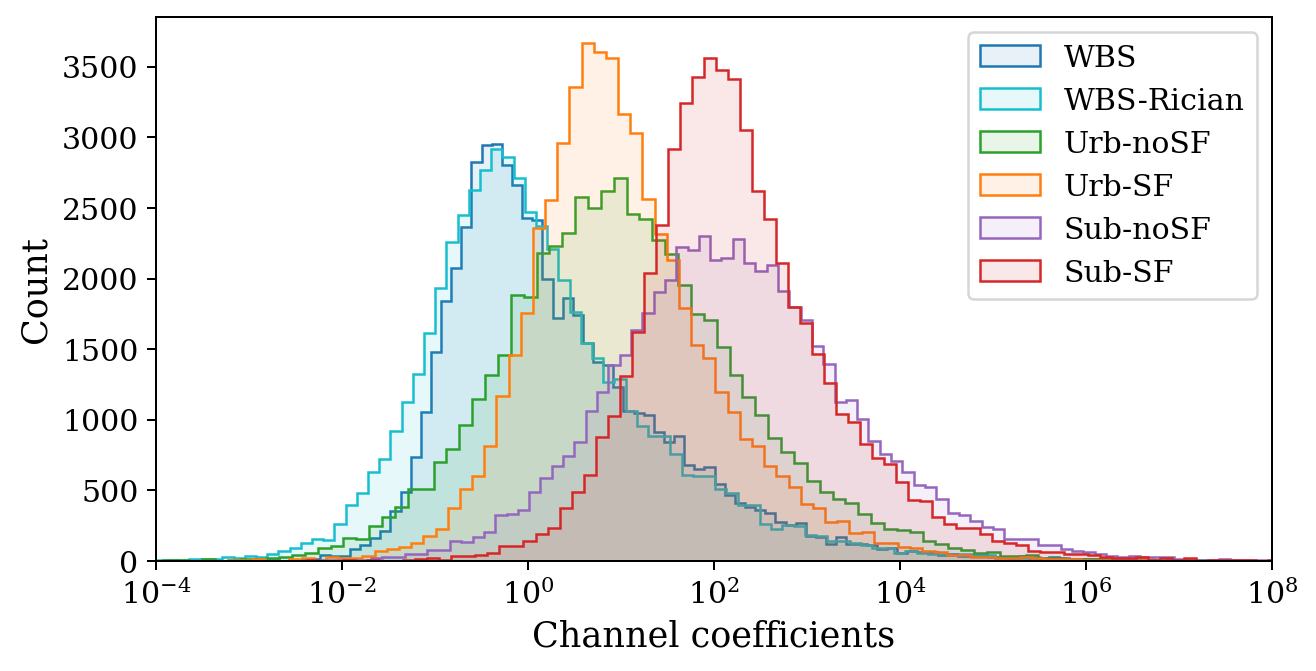}
    \vspace{-6mm}
    \caption{
    Histograms of channel coefficients derived from different PL models with different fading or shadowing effects.
    } \label{fig:data-dist}
\end{figure}

\vspace{-1mm}
In this section, we depart from the {WBS Rayleigh fading channels} and study how robust the allocation methods are to changes in the test channel distribution. 
{We first consider the WBS model with Rician fading, shortened to \mbox{\it WBS-Rician}.} 
Additional test datasets are generated following the Hata-COST231 propagation model~\cite{damosso1999digital}. 
Another four types of PL effects are considered -- urban or suburban environment, with or without the log-normal shadowing -- with abbreviations of \mbox{\it Urb-noSF} (Hata-COST231 urban model without shadowing), \mbox{\it Sub-SF} (Hata-COST231 suburban model with shadowing), \mbox{\it Sub-noSF}, and \mbox{\it Urb-SF}.
We assume $1.9\GHz$ for the carrier frequency and $30\m$ for the BS height; if applicable, the standard deviation of shadowing is $8\dB$.
These distributions are illustrated via histograms of test channel coefficients; see Fig.~\ref{fig:data-dist}. 
More similarity is observed between WBS(-Rayleigh) and WBS-Rician, whereas Urb and Sub distributions are shifted further away from those of WBS.
Since larger coefficients correspond to stronger interference, it generally indicates harder allocation cases -- not necessarily lower achievable objectives but more difficulties in finding the optimal solutions. 

\begin{figure}[t]
	\centering
    \includegraphics[width=\linewidth
    ]{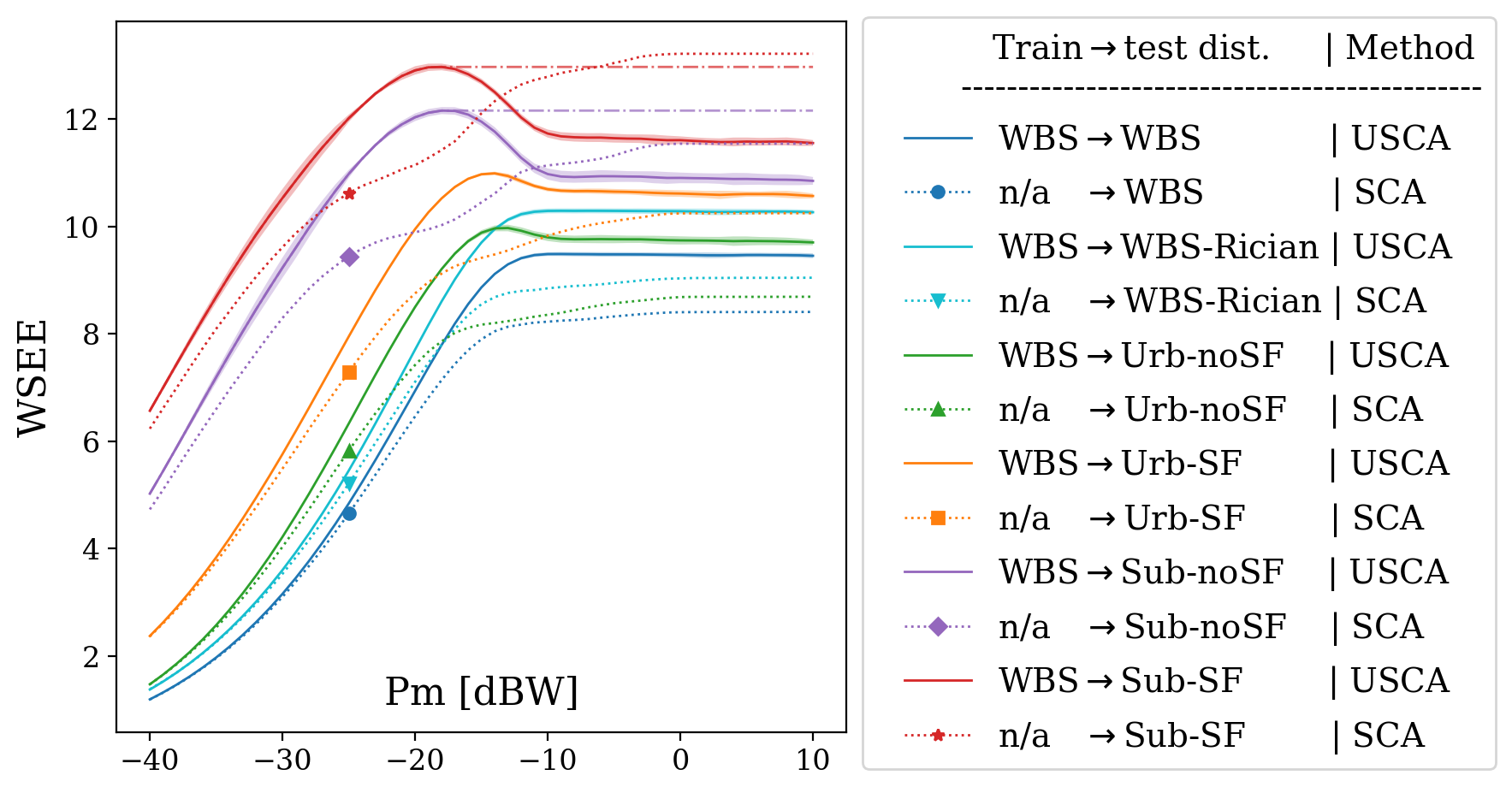}
    \vspace{-5mm}
    \caption{
    Performance of \mbox{GCN-USCA} trained on the WBS channels and tested on different channel distributions compared with that of the classical SCA.} \label{fig:gen-dist}
    \vspace{-4mm}
\end{figure}

Again, for each candidate method, the model being tested here is the exact model from Section~\ref{sec:exp:compare} without any further processing.  
Numerical results are reported in Table~\ref{tab:dist}. 
Additionally, we visualize the curves of average WSEE against the power constraint $P_m$ in the same range as before; see Fig.~\ref{fig:gen-dist}. 
Once more, without any retraining, \mbox{GCN-USCA} achieves the best performance in almost every setting only ranking second in sub-SF behind the classical SCA baseline (excluding from our analysis GCN-USCA$^+$, which will be introduced shortly).

\renewcommand{\tabcolsep}{3.5pt}
\begin{table}[t]
    \caption{{
    Average WSEE values of power allocations on test sets with different channel distributions as a result of different PL propagation models, fading, and shadowing effects.}}\label{tab:dist}
    \vspace{-1mm}
    \centering
    \begin{adjustbox}{max width=\columnwidth}
    \begin{tabular}{r|c c|c c|c c}
        \hline
         PL propa- & \multicolumn{2}{c|}{Wideband Spatial} & \multicolumn{2}{c|}{Hata Urban} & \multicolumn{2}{c}{Hata Suburban}\\\cline{2-7}
         gation model & Rayleigh & Rician  & w/o SF & SF   & w/o SF & SF \\\hline
        GCN-USCA	& {\bf 6.877} & {\bf 7.544} & {\bf 7.583} & {\bf 8.719} & {\bf 10.276} & 11.223 \\
        GCN-USCA-NS	& 6.830 & 7.506 & 7.533 & 8.631 & 10.265 & 11.168 \\
        MLP-USCA	& 6.269 & 6.889 & 6.791 & 7.801 & 8.695 & 9.725 \\
        GCN	& 6.581 & 7.212 & 7.234 & 8.290 & 9.844 & 10.692 \\
        SCA	& 6.203 & 6.751 & 6.712 & 8.118 & 9.821 & {\bf 11.343} \\
        Tr-SCA	& 6.127 & 6.664 & 6.520 & 7.757 & 9.056 & 10.249 \\
        \hdashline
        GCN-USCA$^+$	& 6.883 & 7.548 & 7.681 & 8.878 & 10.858 & 11.855 \\  
        \hline
    \end{tabular}
    \end{adjustbox}
    \vspace{-4mm}
\end{table}

A few observations are in order.
First, the fact that \mbox{GCN-USCA} consistently outperforms \mbox{MLP-USCA} and GCN demonstrates {better generalizability} of our method compared with other learning-based approaches.
Second, notice that the gap between \mbox{Tr-SCA} and SCA is larger for the suburban channels, indicating a more challenging allocation scenario where SCA has slow convergence, which agrees with our intuitions from Fig.~\ref{fig:data-dist}.
This is also consistent with \mbox{GCN-USCA} underperforming in such a setting.
In Fig.~\ref{fig:gen-dist}, if we focus on the solid purple and red curves (and their dotted SCA counterparts), it is clear that \mbox{GCN-USCA} conspicuously outperforms SCA even in the suburban setting for low maximum power constraints {(roughly among $P_m{\,\leq\,}{-}20\dBW$)}.
However, for higher power constraints (i.e., larger feasible space among $P_m{\,\geq\,}{-}10\dBW$), \mbox{GCN-USCA} fails to maintain the monotonicity discussed in Section~\ref{sec:method:usca}, {and thus dragging its overall average performance eventually lower than that of SCA.
Nonetheless, given that} SCA has to be sequentially solved from lower to larger values of $P_m$ (since the solution from one value is used as a warm-start for the next one; see~\cref{ft:sca-init-p}), a fairer comparison would be one where we follow the same procedure for \mbox{GCN-USCA}.
Under such a setting, we can impose monotonicity by simply running \mbox{GCN-USCA} for a given $P_m$ and, if the WSEE decreases with respect to the one obtained for a lower $P_m$, we simply keep using the previous power.
This modified methodology gives rise to the dashed lines in Fig.~\ref{fig:gen-dist} (illustrated for the suburban channels only) and we list the obtained WSEE under \mbox{GCN-USCA}$^+$ in Table~\ref{tab:dist}.
Note that for the WBS models (where training and testing follow the same {or very similar} channel distribution), there is neglectable difference between \mbox{GCN-USCA} and \mbox{GCN-USCA}$^+$ since monotonicity has been successfully imposed already.
On the other hand, for the suburban channels the improvement of \mbox{GCN-USCA}$^+$ is notable, markedly outperforming both \mbox{GCN-USCA} and SCA.

Overall, \mbox{GCN-USCA} has good generalizability to channel models unseen during training.
Even if the mismatch between training and testing data proves to be a major problem, this learning-based method can enjoy effortless improvement by simple transfer learning~\cite{dong2020deep} given some data from the target distribution.
This is further discussed in the next section.

\vspace{-1em}
\subsection{Impact of the training set size in fine-tuning across channel distributions}\label{sec:exp:supervision}
\vspace{-1mm}
We now show that, by simple fine-tuning, our proposed framework, {originally trained on an easy channel distribution, is able to} provide energy-efficient power allocation policies for a different (possibly more difficult) distribution. 
In particular, we fix our attention on transferring \mbox{GCN-USCA} from the WBS training distribution to the Sub-SF test distribution, as the previous section has indicated that this scenario is specially challenging.
The fine-tuning is performed in an end-to-end manner with a universal learning rate of $l_f{\,=\,}5{\times}10^{-5}$. The base model is taken from the 1st fold (out of four CV folds) of the previously trained \mbox{GCN-USCA} models.
Observing that its validation performance saturates with 7 blocks, we abandon the three tailing blocks and set $T^{\prime}{\,=\,}7$ for the fine-tuning experiment.
Primarily, the entire set of \num{1000} Sub-SF channels coupled with \num{51} constraint values {(ranging from ${-}40$ through $10\dBW$)} are used for fine-tuning {in 2-fold CV}; the regularization coefficients $\eta_m$ and $\eta_s$ are set to be \num{0.25} and \num{0} (completely unsupervised), respectively. 
The final average test performance of \mbox{GCN-USCA} is {\num{12.271}}, peaking at \num{13.566} ($P_m{\,=\,}{-}18\dBW$) and then stabilizing around \num{13.373}. 
The stationary point is close to the peak point, indicating that \mbox{GCN-USCA} has {almost restored monotonicity.
As a commensurable baseline}, the WSEE of SCA saturates towards {\num{13.219}} as $P_m$ approaches $0\dBW$.
From these observations, we find that fine-tuning across input distributions can significantly improve allocation performance and renovate monotonicity on the target distribution. 

\begin{figure}[t]
	\centering
    \includegraphics[width=.98\linewidth]{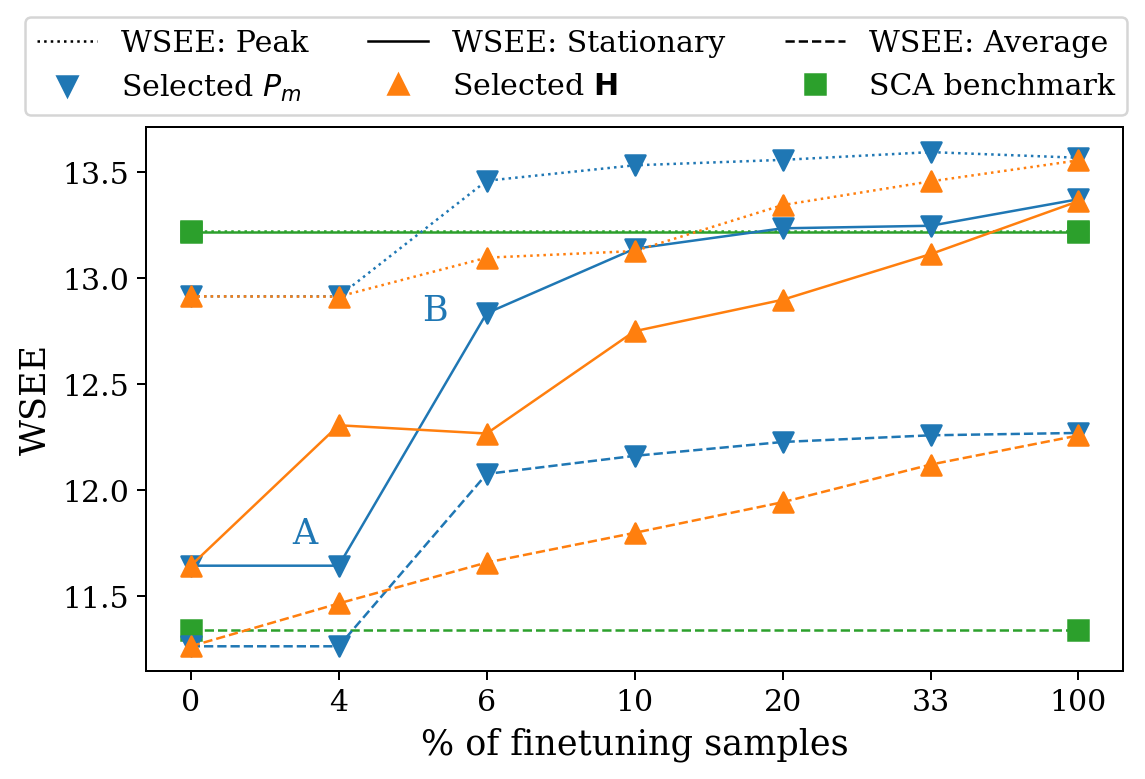}
    \vspace{-2mm}
    \caption{{
   Average, peak, and stationary WSEE values of fine-tuned \mbox{GCN-USCA} power allocations on test set with different amount of training samples by selecting subsets of $\ccalP_m$ or $\ccalH_{s}$.
   The SCA performance (horizontal lines) is annotated for reference.}
    } \label{fig:ft-ph} 
    \vspace{-4mm}
\end{figure}

Next, we proceed to probe the minimal amount of data needed.
Two sets of fine-tuned \mbox{GCN-USCA} performance on the Sub-SF test set are displayed in Fig.~\ref{fig:ft-ph}, resulted from selecting subsets of constraints $P_m$ and channels $\bbH$, respectively.
The WSEE vs $P_m$ curves {(those in Fig.~\ref{fig:wsee:pmax} or Fig.~\ref{fig:gen-dist})} are characterized by three points -- 
i)~the {\it average} denotes the average WSEE value over all test samples; 
ii)~the {\it peak} marks the highest achieved WSEE value within a WSEE-$P_m$ curve, which is an average value over all channel samples at a certain $P_m$;  
iii)~the {\it stationary} represents the point toward which WSEE tends to approach as $P_m$ grows sufficiently large, specifically defined as the mean WSEE value of all samples where $5{\,<\,}P_m{\,\leq\,}10\dBW$. 
Ideally, the stationary point should coincide with the peak point due to monotonicity.  
In the \mbox{GCN-USCA} cases, generally speaking, fine-tuning has indeed {gradually eliminated} the wide gap {between the stationary and peak points} present in the base model, which we elaborate with Fig.~\ref{fig:ft-ph} in the sequel. 

Again, denote the set of 51 possible $P_m$ values by the calligraphic $\ccalP_m$ and the set of 1000 available channel realizations of the target distribution by $\ccalH_{s}$.
First, it is of realistic significance to simply reduce the size of the fine-tuning set, because one would naturally favor a learning scheme that requires fewer samples and epochs. 
To address this need, the most straightforward implementation is to drop certain constraint values. 
We select subsets of $P_m$ by increasing the sampling interval from 1 to {3, 5, 10, 25, and 50$\dBW$}, corresponding to using {100, 33, 20, 10, 6, and 4\% of $\ccalP_m$} (denoted by `Selected $P_m$' in Fig.~\ref{fig:ft-ph}).
We observe a huge jump from point A (\num{4\%}) to B (\num{6\%}).
This corresponds to the additional $P_m$ anchor of ${-}15\dBW$ in the \num{6\%} subset, i.e., $\{{-}40,{-}15,10\}\dBW$ versus $\{{-}40,10\}\dBW$ of the $4\%$ subset.
Beyond point B, the fine-tuned \mbox{GCN-USCA} can easily adapt to the new distribution and outperform SCA {in peak and average.
Please be reminded that as long as GCN-USCA beats SCA in the peak value, the post-processed GCN-USCA$^{+}$ will surpass SCA at every point of $P_m$.
With $20\%$ training samples or more, the stationary performance of the fine-tuned GCN-USCA also catches up with that of SCA, meaning that GCN-USCA can completely outperform SCA even without the post processing while} saving \num{80\%} of training time.

From a practical standpoint, we might lack channel samples when switching to a new environment. 
To accommodate for the lack of channel data, we restrict the amount of accessible $\bbH$ samples by randomly sampling 40, 60, 100, 200, and 333 channel realizations, corresponding to 4, 6, 10, 20, and 33\% of samples in $\ccalH_{s}$ (denoted by `Selected $\bbH$' in Fig.~\ref{fig:ft-ph}).
The coupled $P_m$ values are the entire set of $\ccalP_m$.
A strong dependency on channels is clearly revealed in Fig.~\ref{fig:ft-ph}, where the {$\ccalH_{s}$-subset} performance, though starting off higher than the $\ccalP_m$-subset fine-tuning which starts with very limited constraints, quickly falls behind the $\ccalP_m$-subset performance with the same amount of channel-constraint samples. 
Using only $4\%$ available channel data, the fine-tuned model already gives better {\it average} performance than SCA; however, its {\it peak} performance does not match that of SCA until $20\%$ channel data is employed.

The abovementioned observations indicate that representative maximum power constraints and sufficient channel data are crucial to successful unsupervised transfer learning to a new distribution. 
To sum up, the number of blocks chosen for fine-tuning, the down-sampling of constraint values, and the selection of channel realizations are open questions that one needs to decide depending on specific applications and requirements when deploying the USCA framework in a new environment.


\vspace{-.5em}
\section{Conclusions and Future Work}\label{sec:conclusion}
\vspace{-1mm}
We developed USCA, a novel graph-based learning approach to allocate transmit power for maximizing WSEE in wireless interference networks.
Using algorithm unfolding directed by {the SCA algorithm}, we proposed the modular architecture of USCA that intrinsically incorporates problem-specific modules augmented by trainable components.
To parameterize these components, GCNs are leveraged to make explicit use of the underlying topology of wireless communication networks.
Through extensive analyses and experiments, we have demonstrated that USCA yields solutions that are: 
i)~Near-optimal, since it constantly outperforms the {classical SCA algorithm and other} well established benchmarks;
ii)~Efficient, because the neural architectures, once trained, will consume less time and computation than the classical model-based methods; and 
iii)~Distributed, owing to the decentralized implementation proposed for its modular GCN-based architecture.

A natural direction to explore next would be to obtain energy-efficient power allocation in case of higher degrees of heterogeneity, such as nonuniform system constants, user-specific power or other constraints, and different user weights. 
From a methodological perspective, this work also paves the way for many intriguing avenues for future research in wireless resource allocation not limited to power. 
First, we can dig deeper into understanding the transformed loss landscape of the sum-of-ratios fractional objective by, for example, visualizing the morphology and dynamics of loss function during training or fine-tuning. 
Second, in lieu of the presented post-training fine-tuning, we are interested in ways to inherently promote the model's generalizability, e.g., alternative designs of learning components and distribution diversity in training data. 
Finally, a direction of practical importance is to consider scenarios where channel information is missing, noisy, or even adversarial, which may raise broader problems to address concerning network security.

\begin{spacing}{1.}
\footnotesize\bibliography{ref}
\bibliographystyle{ieeetr}
\end{spacing}

\end{document}